%% file: SpaceOfPCFsArxiv.tex
\documentclass[acmtog]{acmart}
\acmSubmissionID{115}

\usepackage{booktabs} 

\setcitestyle{square}

\usepackage[ruled]{algorithm2e} 

\SetAlFnt{\small}
\SetAlCapFnt{\small}
\SetAlCapNameFnt{\small}
\SetAlCapHSkip{0pt}
\IncMargin{-\parindent}

\acmJournal{TOG}
\acmVolume{0}
\acmNumber{0}
\acmArticle{0}
\acmYear{2019}
\acmMonth{0}

\setcopyright{none}

\acmDOI{0000001.0000001_2}

\setcitestyle{square}

\usepackage{wrapfig}
\usepackage{amssymb,amsmath}
\usepackage{algorithmic}
\usepackage{url}
\usepackage{mathrsfs} 
\usepackage{bm} 

\begin{document}

\title{A Comprehensive Theory and Variational Framework for Anti-aliasing Sampling Patterns} 
\author{A. Cengiz {\"O}ztireli}
\affiliation{%
  \institution{Disney Research Studios}
  \country{Switzerland}}
	
\begin{abstract}

In this paper, we provide a comprehensive theory of anti-aliasing sampling patterns that explains and revises known results, and show how patterns as predicted by the theory can be generated via a variational optimization framework. We start by deriving the exact spectral expression for expected error in reconstructing an image in terms of power spectra of sampling patterns, and analyzing how the shape of power spectra is related to anti-aliasing properties. Based on this analysis, we then formulate the problem of generating anti-aliasing sampling patterns as constrained variational optimization on power spectra. This allows us to not rely on any parametric form, and thus explore the whole space of realizable spectra. We show that the resulting optimized sampling patterns lead to reconstructions with less visible aliasing artifacts, while keeping low frequencies as clean as possible.

\end{abstract}

%
%

\ccsdesc[500]{Computing methodologies~Antialiasing}
\ccsdesc[300]{Computing methodologies~Rendering}
\ccsdesc[300]{Computing methodologies~Image processing}
%
%

\keywords{Sampling, anti-aliasing, stochastic point processes, image processing, rendering}
 
\thanks{}
	
\begin{teaserfigure}
   \includegraphics[width=0.99\linewidth]{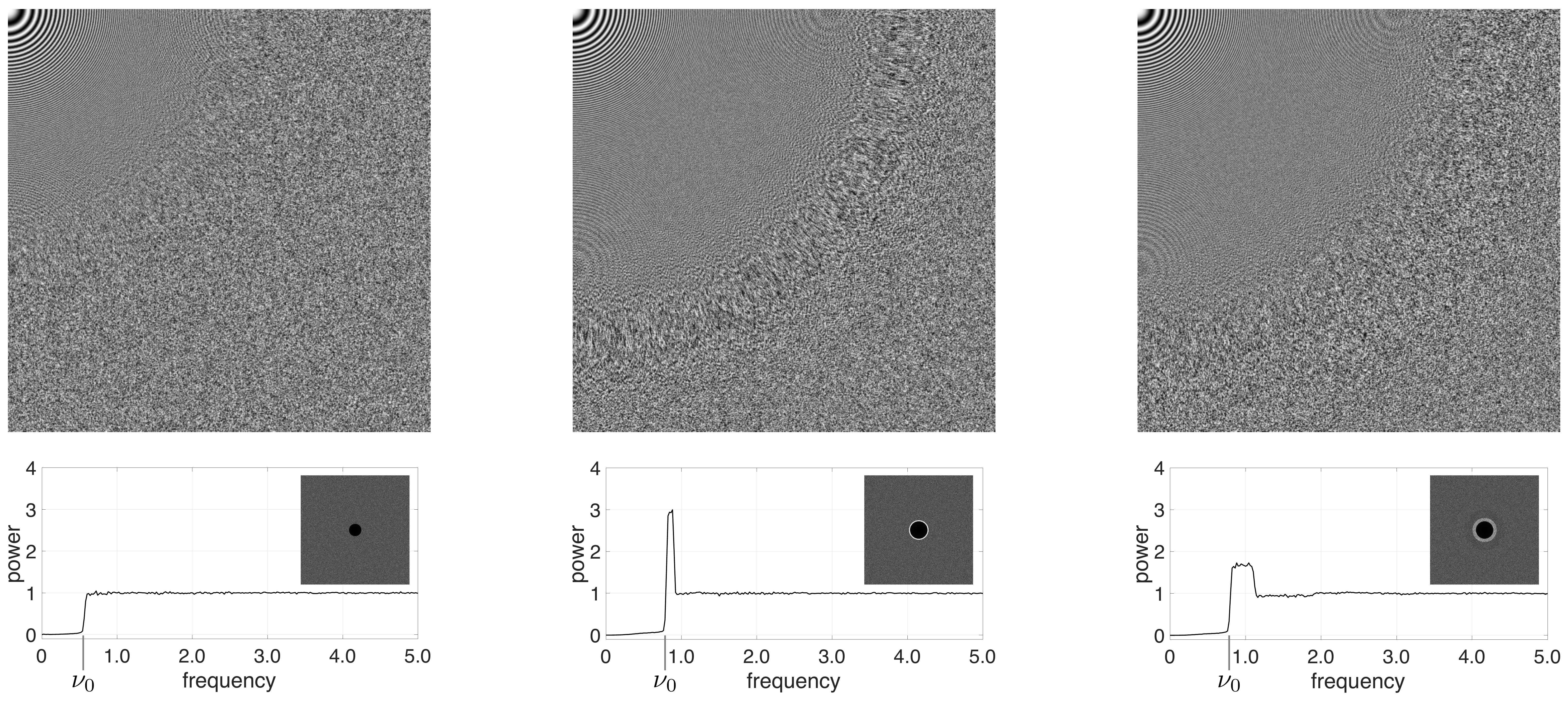}
   \caption{Anti-aliasing patterns, such as the step blue noise on the left, can generate images with clean low frequency content, and map higher frequencies to incoherent noise. The range of clean low frequencies (determined by $\nu_0$ here) can be increased at the cost of introducing coherent colored noise for higher frequencies (middle, stair blue noise~\protect\cite{Kailkhura16Stair}). Our approach (right) can generate sampling patterns that introduce minimal aliasing, while keeping the same range of clean low frequencies (top: reconstructed zone plate images, bottom: 1D power spectra, insets: 2D power spectra).}
   \label{fig:teaser}
\end{teaserfigure}

\maketitle

\input{TexFiles/Introduction}

\input{TexFiles/RelatedWork}
\input{TexFiles/Method}

\input{TexFiles/Results}

\input{TexFiles/Conclusions}

\appendix
\input{TexFiles/Appendix}

\bibliographystyle{ACM-Reference-Format}
\bibliography{SpaceOfPCFsArxiv}

\end{document}

%% file: TexFiles/Introduction.tex
\section{Introduction}

Sampling patterns are fundamental for many applications in computer graphics such as imaging, rendering, geometry sampling, natural distribution modeling, among others. They are of particular importance for reconstructing images from samples. Most real-world or synthesized images are not band-limited, i.e. they contain frequencies higher than those that can be represented with a finite number of samples, inevitably leading to aliasing. The challenge is avoiding aliasing artifacts that show up as secondary structures that are not present in the original image, while having the lower frequency content cleanly reconstructed. An ideal anti-aliasing sampling pattern thus preserves lower frequencies by introducing minimal noise, and maps all higher frequencies that cannot be represented with the sample budget to incoherent noise, instead of visible artifacts~\cite{Dippe85Antialiasing,Cook86Stochastic}. 

A family of patterns proposed to approximate these properties are \emph{blue noise} patterns~\cite{Ulichney88Dithering}. A zone plate image sampled with such a pattern, and the 2D and 1D power spectrum $P(\nu)$ of the corresponding sampling pattern are depicted in Figure~\ref{fig:teaser}, left. The goal of blue noise patterns is to keep $\nu_0$ as large as possible, while minimizing deviations from $1$ for higher frequencies, with the intuition that the former will ensure clean low frequencies while the latter will lead to minimal aliasing. However, a perfectly flat power spectrum with $P(\nu) = 1$ for  $\nu > \nu_0$ is only possible for quite low values of $\nu_0$, and in general only random sampling can have a constant spectrum of $P(\nu) = 1$ for all $\nu$. Low $\nu_0$ values lead to noisy low frequency content for sampled images, as visible in the limited clean region in the upper left corner of the zone plate image in Figure~\ref{fig:teaser}, left. 

Many techniques have been proposed to generate sampling patterns for a larger range of cleanly represented frequencies, while avoiding aliasing artifacts as much as possible. Initial efforts focused on designing algorithms that impose constraints on certain geometric properties of the sampling patterns, such as the classical dart throwing algorithm~\cite{Cook86Stochastic} and its variations, where sampling points are randomly distributed with a minimum distance between each pair. More recent techniques assume provided spectra or related statistics, and optimize locations of sampling points such that the resulting distributions have the given statistics~\cite{Zhou12Point,Oztireli12Analysis,Heck13Blue,Wachtel14Fast,Ahmed15AA,Kailkhura16Stair}. This approach provides generic algorithms that can generate sampling patterns with any given characteristics such as a power spectrum.

The challenge, however, is how to specify useful shapes for power spectra in the limited space of realizable spectra~\cite{Uche06Ontherealizability}. Recent works have focused on generating realizable spectra with certain properties beneficial for anti-aliasing~\cite{Heck13Blue,Kailkhura16Stair}. These methods assume a parametric form for power spectra, and search in the parameter space to have the least energy in the low frequency region, and a flat high frequency region bounded from above. Such a sampling pattern generated by a state-of-the-art technique~\cite{Kailkhura16Stair} is shown in Figure~\ref{fig:teaser}, middle. The $\nu_0$ is significantly larger than that of step blue noise (left), which manifests itself as a larger range of clean low frequencies in the reconstructed zone plate image. However, this comes at the cost of a flat peak in the power spectrum, introducing artifacts in the zone plate image for middle frequencies. In general, assuming a given parametric form limits power spectra, leading to sub-optimal anti-aliasing properties.

In this paper, we introduce a comprehensive theoretical framework for anti-aliasing that leads to a variational approach to compute power spectra with optimized characteristics with respect to their anti-aliasing properties. In order to formulate the corresponding optimization problem, we first prove an analytic form for the spectrum of expected error introduced by sampling in terms of the power spectrum of the function to be represented and that of the sampling pattern. Based on this formula for error, we show how existing patterns improve anti-aliasing, and provide new theoretical results and insights. These are then translated into constraints and energies for formulating a constrained variational optimization on power spectra of point patterns. We show that careful selection of constraints and energies to minimize leads to sampling patterns with improved anti-aliasing properties. A sampling pattern generated by the proposed technique is shown in Figure~\ref{fig:teaser}, right. We get the same $\nu_0$ and thus range of noise-free lower frequencies as for the result of Kailkhura et al.~\shortcite{Kailkhura16Stair}, while still mapping all higher frequencies to almost white noise, as can be seen in the zone plate test image. The resulting power spectrum arises from our formulation of the optimization, without explicitly specifying its form.

In summary, we have the following main contributions:
\begin{itemize}
  \item  A theory of anti-aliasing with exact expressions for expected error spectrum. This allows us to analyze desirable properties for power spectra of point patterns for anti-aliasing.
	\item A new formulation of the problem of generating realizable spectral or spatial characteristics of point patterns based on variational optimization. We study different measures for optimality of sampling patterns, and show that there is a very rich family of realizable characteristics with desirable properties.
	\item Sampling patterns optimized for anti-aliasing with practical improvements over state-of-the-art patterns.
\end{itemize}


%% file: TexFiles/RelatedWork.tex
\section{Related Work}

Aliasing is a fundamental problem when reconstructing or synthesizing images with samples, as the images are typically not band-limited and we always have a finite budget of samples. It is well-known that regular sampling leads to structured aliasing, which introduces visually distracting extra structures. A main observation is that by injecting randomness into point distributions while satisfying certain properties, structured artifacts can be replaced with noise that is potentially visually less distractive~\cite{Dippe85Antialiasing,Cook86Stochastic}. With such random distributions, it is important that high frequencies that cannot be represented with the sample budget are mapped to as incoherent as possible noise, ideally white noise to avoid any extra patterns in the reconstructed image, while keeping the important low frequency content clean. 

Such sampling patterns are typically called \emph{blue noise} in computer graphics. Blue noise patterns are characterized by a low energy power spectrum $P(\nu)$ for $\nu < \nu_0$, and a flat spectrum with $P(\nu) \approx 1$ for $\nu > \nu_0$~\cite{Yellott83Spectral,Mitchell91Spectrally}. Many methods have been proposed to generate point patterns with power spectra that exhibit variations of such properties. Earlier methods propose algorithms that impose certain constraints on the generated random point distributions. Dart throwing~\cite{Cook86Stochastic} (also known as simple sequential inhibition and random sequential adsorption~\cite{Illian08Statistical}) generates distributions where points are randomly placed in space with the constraint that no two points are closer to each other than a certain distance. This algorithm and the resulting distributions have been widely used and extended in many ways in the last decades (e.g.~\cite{Dunbar06Aspatial,Bridson07Fast,Wei08Parallel,Wei10Multi,Ebeida12ASimple,Ebeida14KdDarts,Yuksel15Sample,Kailkhura16Theoretical}). Other works have investigated utilizing alternative algorithms for improved characteristics for certain applications~\cite{Kopf06Recursive,Ostromoukhov07Sampling,Illian08Statistical,Balzer09Capacity,Schmaltz10Electrostatic,Fattal11Blue,Schlomer11Farthest,Xu11Capacity,Chen12Variational,deGoes12Blue,Jiang15Blue}. The resulting point distributions are then analyzed by computing characteristics such as power spectrum, or statistics from stochastic point processes~\cite{Mitchell87Generating,Dutre08AComparison,Oztireli12Analysis,Heck13Blue}, to understand their utility in practice.

A main limitation of the mentioned works for point pattern generation, however, is that the algorithm dictates the characteristics of the generated point patterns. Instead, a recent body of works propose to generate point distributions with statistics matching given ones~\cite{Zhou12Point,Oztireli12Analysis,Heck13Blue,Wachtel14Fast,Ahmed15AA,Kailkhura16Stair}. Once a statistic, such as power spectrum, is defined, these methods run a routine to place sampling points such that the final configuration leads to the desired form for the statistic. With this approach, Heck et al.~\shortcite{Heck13Blue} could generate point distributions with the step blue noise spectrum for the first time (Figure~\ref{fig:teaser}, left). However, they have also observed that such a form for the spectrum is only possible for quite low values of $\nu_0$, leading to noisy lower frequencies for sampled images. In general, the sub-space of realizable power spectra is restricted, with the necessary conditions that both power spectrum and pair correlation function, which is related to power spectrum with a spectral transform, should be non-negative~\cite{Uche06Ontherealizability}. Hence, a fundamental challenge is defining realizable forms for power spectra with desirable properties.

This challenge has been addressed by defining parametric forms for power spectrum in recent works~\cite{Heck13Blue,Kailkhura16Stair}. The idea is then to search over the free parameters to get realizable power spectra with anti-aliasing properties. Heck et al.~\shortcite{Heck13Blue} define the \emph{single-peak blue noise}, where a Gaussian is placed at around $\nu_0$ to trade off energy for $\nu < \nu_0$ against the maximum value $m$ of the power spectrum. The standard deviation, and magnitude of the Gaussian can then be altered to get realizable power spectra. Kailkhura et al.~\shortcite{Kailkhura16Stair} have recently proposed a new parametrized family, \emph{stair blue noise}, where the peak is replaced with a raised flat region of a certain width starting at $\nu_0$, as in Figure~\ref{fig:teaser}, middle. The free parameters in this case are $\nu_0$, and the width and height of the raised flat region. By a guided search over these parameters, spectra with a lower $m$ than single-peak blue noise can be obtained. However, for both methods, due to the assumed parametric forms, the families of power spectra considered are rather limited, and the exact effect of the parameters on aliasing is not clear.

In contrast, we do not assume any particular form for power spectra and instead formulate the problem of generating desirable and realizable spectra as constraint optimization with a variational formulation. This formulation rests on a theoretical analysis of properties of power spectra for anti-aliasing. Such an analysis has not been possible before due to the lack of an exact relation between error and power spectra of point patterns. After deriving this relation, we revise common properties of desirable power spectra. By formulating such anti-aliasing specific properties in addition to realizability conditions as constraints and energies, we then get diverse families of power spectra by variational optimization. This framework thus allows us to obtain optimal point patterns with respect to the imposed properties, e.g. for a given $\nu_0$ and perfectly zero energy for $\nu < \nu_0$, we can get the minimum possible $m$, up to numerical accuracy. The resulting point patterns lead to image reconstructions with less artifacts for high frequencies, and cleaner low frequency content.

%


%% file: TexFiles/Method.tex



\section{Background and Preliminaries}
\label{sec:background}

We utilize the theory of stochastic point processes~\cite{Moller03Statistical,Illian08Statistical} to understand anti-aliasing properties of point patterns. Stochastic point processes provide a principled approach for analyzing point patterns~\cite{Moller03Statistical,Illian08Statistical}. A point process is defined as the generating process for multiple point distributions sharing certain characteristics. Hence, each distribution can be considered as a realization of an underlying point process (we use the term \emph{point pattern} for families of point distributions sharing characteristics ).

We can explain a point process with joint probabilities of having points at certain locations in space. Such probabilities are expressed in terms of \emph{product densities}. For our application of point patterns with optimal power spectra for anti-aliasing, it is sufficient to consider first and second order product densities, as they uniquely determine the power spectrum of a point process. First order product density is given by $\varrho^{(1)}(\mathbf{x}) d\mathbf{x} = p(\mathbf{x})$, where $p(\mathbf{x})$ is the probability of having a point generated by the point process $\mathcal{P}$ in the set $d\mathbf{x}$ of infinitesimal volume, and intuitively measures expected number of points around $\mathbf{x}$, i.e. local density. Similarly, second order product density $\varrho^{(2)}$ is defined in terms of the joint probability $p(\mathbf{x},\mathbf{y})$ of having points $\mathbf{x}$ and $\mathbf{y}$ in the sets $d\mathbf{x}$ and $d\mathbf{y}$ simultaneously, $\varrho^{(2)}(\mathbf{x},\mathbf{y}) d\mathbf{x} d\mathbf{y} = p(\mathbf{x},\mathbf{y})$. It describes how points are arranged in space, and is fundamentally related to the power spectrum of $\mathcal{P}$.


As in previous works~\cite{Dippe85Antialiasing,Heck13Blue,Kailkhura16Stair} on anti-aliasing, we will assume that no information is given on the function to be represented, and hence consider unadaptive point patterns. These patterns are generated by stationary and isotropic point processes, where the characteristics of the generated point distributions are translation invariant, or translation and rotation invariant, respectively~\cite{Oztireli12Analysis}. For both cases, $\varrho^{(1)}$ reduces to a constant number, $\lambda$, which measures the expected number of points in any given volume, $\lambda = \frac{\mathbb{E}_{\mathcal{P}} [n(\mathcal{V})]}{|\mathcal{V}|}$, where $\mathbb{E}_{\mathcal{P}}$ denotes expectation over different distributions generated by the point process $\mathcal{P}$, $n(\mathcal{V})$ is the random number of points that fall into the set $\mathcal{V}$, and $|\mathcal{V}|$ is its volume. For stationary point processes, second order product density becomes a function of the difference vector between point locations $\varrho^{(2)}(\mathbf{x},\mathbf{y}) = \varrho^{(2)}(\mathbf{x} - \mathbf{y})$, which can be expressed in terms of the normalized \emph{pair correlation function (PCF)} $g$ as $\varrho^{(2)}(\mathbf{x} - \mathbf{y}) = \lambda^2 g(\mathbf{x} - \mathbf{y})$. For isotropic point processes, PCF further simplifies and becomes a function of the distance between point locations $g(\mathbf{x} - \mathbf{y}) = g(\Vert \mathbf{x} - \mathbf{y} \Vert)$. Below we will first consider stationary point processes and the associated derivations, which we will specialize to isotropic processes in the next sections.

\subsubsection*{PCF as a Distribution}

The intuition behind PCF is that it can be estimated as a probability distribution of difference vectors (for stationary point processes), or distances (for isotropic point processes) between points. This is possible due to the fundamental Campbell's theorem~\cite{Illian08Statistical} that relates sums of functions at sample points to integrals of those functions. For simplicity of the expressions, we assume a toroidal domain $\mathcal{V}$ with unit volume as the sampled domain (e.g. the image plane). Utilizing Campbell's theorem, it is possible to derive the following expression for PCF of stationary processes (Appendix~\ref{sec:appDerivationOfPCF})
\begin{equation}
\begin{aligned}
g(\mathbf{r}) = \frac{1}{\lambda^2} \mathbb{E}_{\mathcal{P}} \biggl[ \sum_{j \neq k} \delta(\mathbf{r} - \mathbf{r}_{jk}) \biggl],
\label{eq:pcfNdDefinition}
\end{aligned}
\end{equation}
where $\delta$ is the Dirac delta, and we defined $\mathbf{r} = \mathbf{x} - \mathbf{y}$, $\mathbf{r}_{jk} = \mathbf{x}_j - \mathbf{x}_k$. Note that the $\mathbf{x}_k$'s are from a particular distribution generated by the point process $\mathcal{P}$, and the expectation is over all such distributions. This expression clearly shows that PCF is simply a normalized distribution of difference vectors $\mathbf{r}_{jk}$.

\subsubsection*{Power Spectrum and PCF}

Power spectrum of a point process is defined in terms of the Fourier transform $S(\bm{\nu}) =  \mathscr{F} [s(\mathbf{x})](\bm{\nu})$ of the function $s(\mathbf{x}) = \sum_j {\delta (\mathbf{x} - \mathbf{x}_j)}$ as follows~\cite{Heck13Blue}
\begin{equation}
\begin{aligned}
P(\bm{\nu}) = \frac{1}{\lambda} \mathbb{E}_{\mathcal{P}} \left[ \overline{S(\bm{\nu})}S(\bm{\nu}) \right] = \frac{1}{\lambda} \mathbb{E}_{\mathcal{P}} \biggl[ \sum_{jk} e^{-2\pi i \bm{\nu}^T \mathbf{r}_{jk}}  \biggl],
\label{eq:powerSpectrumNdDefinition}
\end{aligned}
\end{equation}
with $\overline{(\cdot)}$ denoting complex conjugate. Power spectrum $P$ thus lacks the phase of the Fourier transform and hence is translation invariant, only depending on the difference vectors $\mathbf{r}_{jk}$. Equations~\ref{eq:pcfNdDefinition} and~\ref{eq:powerSpectrumNdDefinition} suggest that $P$ and $g$ are related by a Fourier transform. Indeed, denoting the Fourier transform of $g$ with $G$, it is possible to derive the following relation between them (Appendix~\ref{sec:appRelationPCFAndPS})
\begin{equation}
\begin{aligned}
P(\bm{\nu}) = \lambda G(\bm{\nu}) + 1.
\label{eq:powerSpectrumAndPCF}
\end{aligned}
\end{equation}
In order to state properties of power spectra for anti-aliasing, we will work with a slightly modified form of Equation~\ref{eq:powerSpectrumAndPCF}, where we rewrite the relation between $P$ and $g$ in terms a function $u$ we define, and its Fourier transform $U$, as follows
\begin{equation}
\begin{aligned}
g(\mathbf{r}) = u(\mathbf{r}) / \lambda + 1, \hspace{5mm} P(\bm{\nu}) = U(\bm{\nu}) + 1 + \lambda \delta(\bm{\nu}).
\label{eq:powerSpectrumAndPCFInTermsOfU}
\end{aligned}
\end{equation}


\subsubsection*{Conditions for Realizable Power Spectra}

Power spectrum is non-negative by definition (Equation~\ref{eq:powerSpectrumNdDefinition}), and this is also true for PCF as a distribution of difference vectors (Equation~\ref{eq:pcfNdDefinition}). Hence, two necessary conditions for a valid power spectrum of a point process are
\begin{equation}
\begin{aligned}
g(\mathbf{r}) \geq 0, \hspace{5mm} P(\bm{\nu}) \geq 0.
\label{eq:PSRealizeConditions}
\end{aligned}
\end{equation}
It is still an open question whether these are also sufficient conditions, but no counterexamples have been shown in statistics and physics (e.g.~\cite{Torquato02Controlling}), and these conditions have been successfully used to generate realizable power spectra in previous works~\cite{Heck13Blue,Kailkhura16Stair}.



\subsubsection*{Error in Sampling a Function}

Sampling a function $t$ with a point distribution generated by a point process $\mathcal{P}$ can be written as $s(\mathbf{x})t(\mathbf{x})$ in the spatial domain, or as $[S * T](\bm{\nu})$ in the frequency domain, where $T$ is the Fourier transform of $t$, and $*$ denotes convolution. This sampled representation introduces an error. In order to analyze magnitude and distribution of error, the expected power spectrum of error needs to be computed~\cite{Dippe85Antialiasing,Heck13Blue}
\begin{equation}
\begin{aligned}
E(\bm{\nu}) = \mathbb{E}_{\mathcal{P}} \left[ \lvert [S * T](\bm{\nu}) / \lambda - T(\bm{\nu})  \rvert^{2} \right],
\label{eq:errorInSampling}
\end{aligned}
\end{equation}
where $| \cdot |$ denotes magnitude of a complex number, and the sampled representation is divided by $\lambda$ to normalize the energy of the sampled function~\cite{Heck13Blue}, or equivalently to have an unbiased estimator since $\mathbb{E}_{\mathcal{P}} [s(\mathbf{x})t(\mathbf{x}) / \lambda] = t(\mathbf{x})$ (by applying Equation~\ref{eq:campbellOneVariableStationary} in Appendix~\ref{sec:appDerivationOfPCF}).

We need to relate the error $E(\bm{\nu})$ to $P(\bm{\nu})$ in order to derive desired properties for this statistic $P(\bm{\nu})$, which we elaborate on in the next section.

\begin{figure*}[t!]
  \centering
  \includegraphics[width=0.99\linewidth]{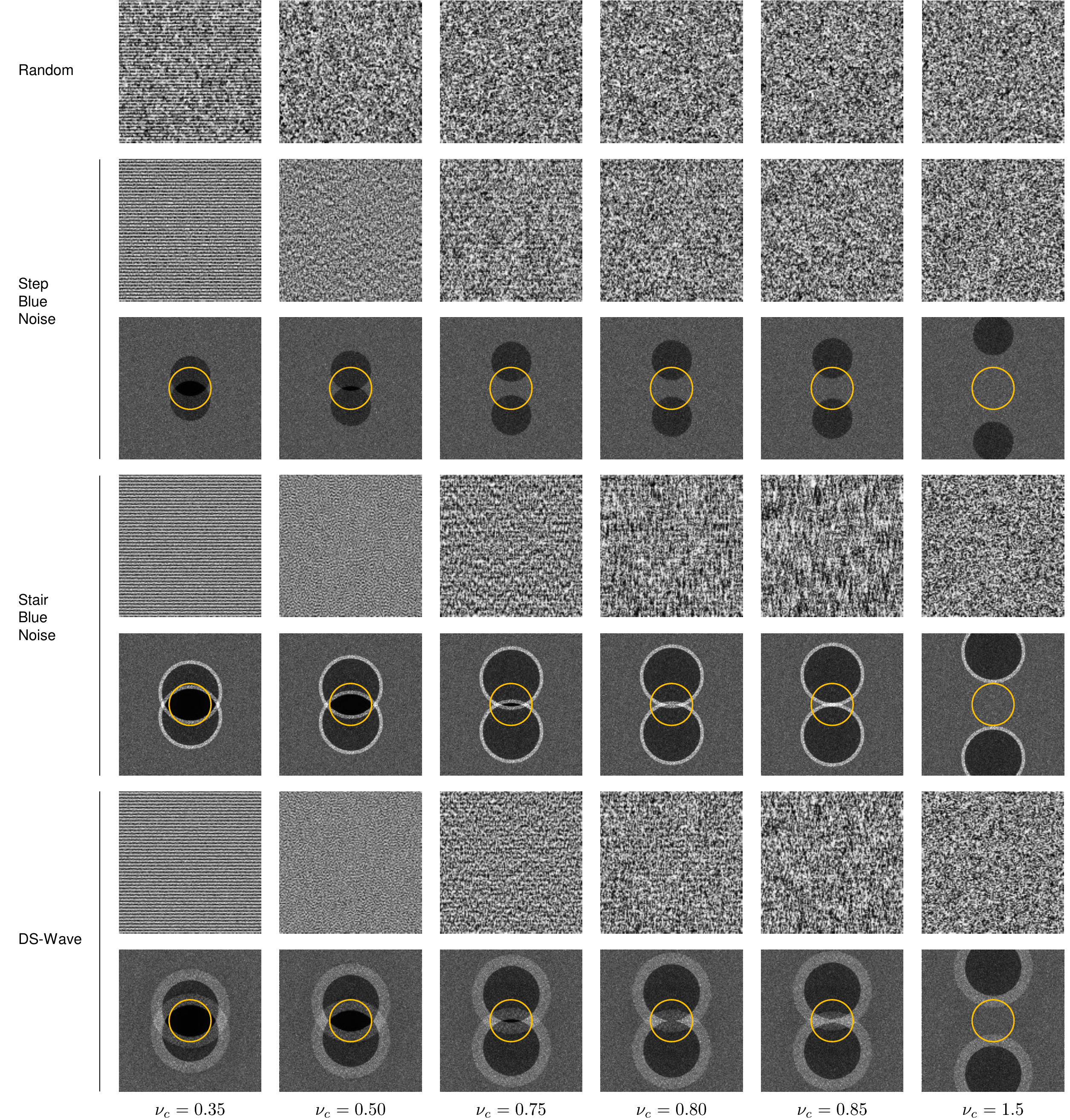}
 \caption{Colored noise leads to visible secondary structures that are not present in original images. The function $\cos ( 2 \pi \sqrt{\lambda} \nu_c y )$ is sampled with different sampling patterns (with $\lambda = 128^2$, and one sample per pixel). We show the resulting images as well as the error $E(\bm{\nu})$, with the orange circle marking the region of representable frequencies after reconstruction and resampling to the pixel grid. Stair blue noise reduces noise levels as compared to step blue noise for lower $\nu_c = 0.35, 0.50$, but leads to aliasing artifacts for higher $\nu_c = 0.75, 0.80, 0.85$ due to the fluctuations it introduces to $E(\bm{\nu})$ for the representable low frequencies within the orange circles. The proposed ds-wave sampling results in less or equivalent noise levels for all $\nu_c$ as compared to other patterns, and less aliasing for higher $\nu_c$ as compared to the state-of-the-art stair blue noise~\protect~\cite{Kailkhura16Stair} (both patterns are with $\nu_0 = 0.8$) due to lower peaks in $E(\bm{\nu})$. For very high $\nu_c = 1.5$, all patterns have noise levels similar to random sampling.
  \label{fig:CosSampling}}\vspace{-0mm}
\end{figure*}

\section{Theoretical Analysis of Sampling Error}
\label{sec:theoreticalAnalysisOfError}

The error spectrum $E(\bm{\nu})$ provides how much error we get at each frequency. In general, we need to have $E(\bm{\nu})$ as low as possible at each $\bm{\nu}$, and especially for low $\bm{\nu}$. For anti-aliasing, we need to additionally have an as uniform as possible $E(\bm{\nu})$ to get incoherent noise instead of colored noise~\cite{Dippe85Antialiasing,Heck13Blue}. It is, however, not clear how these are exactly related to the shape of the power spectrum $P(\bm{\nu})$ of a sampling pattern. We need this relation to be able to formulate the constraints and energies for our variational formulation of optimized sampling patterns for anti-aliasing.

\subsection{Spectra of Error and Sampling Patterns}
\label{sec:errorAndSpectrum}

So far, relating $E(\bm{\nu})$ to $P(\bm{\nu})$ has only been possible for a constant function $t(\mathbf{x}) = c$~\cite{Dippe85Antialiasing}, or upper bounds could be derived for a sinusoidal wave~\cite{Heck13Blue}. In this section, we show that it is possible to derive an exact relation between $E(\bm{\nu})$ and $P(\bm{\nu})$ for an arbitrary $t(\mathbf{x})$, by utilizing the theory of point processes. This leads to theoretical justifications of criteria used for $P(\bm{\nu})$ in the literature, and to novel theoretical results and insights.

We start by expanding the expression for the error spectrum in Equation~\ref{eq:errorInSampling} (we drop $\bm{\nu}$ for brevity)
\begin{equation}
\begin{aligned}
E &= \frac{1}{\lambda^2}\mathbb{E}_{\mathcal{P}} \left[ |S * T|^2 \right] + \mathbb{E}_{\mathcal{P}} \left[ |T|^2 \right] \\
&- \frac{1}{\lambda} \mathbb{E}_{\mathcal{P}} \left[ \overline{ \left( S * T \right) } T \right] 
 - \frac{1}{\lambda} \mathbb{E}_{\mathcal{P}} \left[ \left( S * T \right) \overline{T} \right] \\
&= \frac{1}{\lambda^2}\mathbb{E}_{\mathcal{P}} \left[ |S * T|^2 \right] + |T|^2
  - \frac{2}{\lambda} \Re \left\{ \left( \mathbb{E}_{\mathcal{P}} \left[ S\right] * T \right) \overline{T} \right\},
\label{eq:errorSpectrumExpansion}
\end{aligned}
\end{equation}
where $\Re(\cdot)$ gives the real part of a complex number. The critical part of the proof is deriving the forms of these expected values. We show in Appendix~\ref{sec:derivationPSError} that this can be achieved by starting from Campbell's theorem (as defined in Appendix~\ref{sec:appDerivationOfPCF}). The final form of the power spectrum of error is then
\begin{equation}
\boxed{ E(\bm{\nu}) = \frac{1}{\lambda} \left[ P_t * (U + 1) \right] (\bm{\nu}) . }
\label{eq:errorSpectrumAndP}
\end{equation}
Here, $P_t = \lvert T \rvert^2$ is the power spectrum of the function $t$.
\subsubsection*{Remarks} 
This expression immediately reveals several interesting properties of error when sampling a function.

\begin{itemize}
	

\item The error is independent of the phase of $T(\bm{\nu})$. This is expected as the sampling patterns considered are translation invariant.

\item It implies that error \emph{can} decrease as $O({\lambda}^{-1})$ for any function $t$, as observed for a sinusoidal wave previously~\cite{Heck13Blue}. However, at the same time, the difference vectors $\mathbf{r}_{jk}$ become smaller for higher number of points, leading to a compression of the domain of $g(\mathbf{r})$ (Equation~\ref{eq:pcfNdDefinition}), and hence an expansion of that of $U(\bm{\nu})$, as they are related via a Fourier transform. Thus, the final convergence rate depends on $P_t(\bm{\nu})$ and $U(\bm{\nu})$. 

\item The only pattern that gives a constant spectrum is random sampling (Poisson point process) with $U(\bm{\nu}) = 0$. In this case, we get $E(\bm{\nu}) = \frac{1}{\lambda} \int_{-\infty}^{\infty} P_t(\bm{\nu}) d \bm{\nu}$, which leads to equally noisy frequencies and hence perfectly incoherent white noise. 

\end{itemize}

In practice, a function (e.g. image) sampled with an anti-aliasing point pattern is then resampled to a regular grid after low-pass filtering. This can be written as $s_{REG} ( k * (s t) )$ (dropping $\mathbf{x}$ for brevity), where $k$ is a low-pass filter such as Gaussian, and the points in $s_{REG}$ are regularly distributed on a grid of e.g. pixel centers. This resampled function has the Fourier transform $S_{REG} * (K (S * T))$ with $S_{REG}$ and $K$ the Fourier transforms of $s_{REG}$ and $k$, respectively. As $S_{REG}$ is an impulse train, the result of this convolution is repeating the same function, assuming $K$ avoids any overlap between aliases. Thus, only the central part around zero frequency, cut out by the filter $K$, is relevant. The expected error (Equation~\ref{eq:errorInSampling}) then becomes 
\begin{equation}
\boxed{ \mathbb{E}_{\mathcal{P}} \left[ \lvert K [S * T] / \lambda - K T  \rvert^{2} \right] = \lvert K \rvert^2 E . }
\label{eq:errorSpectrumAndPFiltered}
\end{equation}
Hence, we can consider the low frequency region of $E$ implied by $K$ for most practical applications.

\subsubsection*{Relation to integration}
In this work, we are interested in error when representing a function with samples. This is fundamentally different than the error introduced by numerically integrating a function by summing the sample values. However, we can think of the sampling, filtering, and resampling of an image as performing \emph{local} integration around each pixel center. This becomes clear if we explicitly write the process to compute $s_{REG} ( k * (s t) )$. First, we can write $s(\mathbf{x}) t(\mathbf{x}) = \sum_j \delta(\mathbf{x} - \mathbf{x}_j) t(\mathbf{x}_j)$. Convolving this with $k$, we get $\sum_j k (\mathbf{x} - \mathbf{x}_j) t(\mathbf{x}_j)$. Finally, evaluating it at each pixel center $\mathbf{c}_k$, we get $\sum_j k (\mathbf{c}_k - \mathbf{x}_j) t(\mathbf{x}_j)$. Normalized by $\lambda$, this can be considered as a numerical approximation of the integral $\frac{1}{|\mathcal{V}|}\int_{\mathcal{V}} k (\mathbf{c}_k - \mathbf{x}) t(\mathbf{x}) d \mathbf{x}$ ($|\mathcal{V}$| is the volume of $\mathcal{V}$). To understand aliasing, we need to analyze the \emph{distribution} of these errors of integral estimates at all pixels. Indeed, we are interested in the spectrum of error that encodes this distribution. This is in contrast with analyzing error in a single integral estimate. 

Equation~\ref{eq:errorSpectrumAndP} further reveals an interesting relation with integration error. The DC component of sampling error given by $E(\mathbf{0}) = \frac{1}{\lambda} \int_{-\infty}^{\infty}  P_t(\bm{\nu}) (U(\bm{\nu}) + 1)  d \bm{\nu}$ is exactly the variance of the numerical estimator $\frac{1}{\lambda} \sum_j t(\mathbf{x}_j)$ for the integral $\frac{1}{| \mathcal{V}|} \int_{\mathcal{V}} t(\mathbf{x}) d \mathbf{x}$~\cite{Pilleboue15Variance,Oztireli16Integration}. For the stationary point processes we consider, bias vanishes and hence this variance is equal to the expected error of the numerical integral estimator~\cite{Oztireli16Integration}. 

\subsection{Analysis of Anti-aliasing Properties}
\label{sec:analysisOfAntialiasingProperties}

The derived relation between $E(\bm{\nu})$ and $P(\bm{\nu})$ allows us to perform a theoretical analysis of error in terms of the characteristics of the power spectrum. There are established characteristics for the power spectra $P(\bm{\nu})$ of anti-aliasing point patterns in the literature. These follow certain intuitions and have indeed been effective in practice. However, how such characteristics exactly affect aliasing, and how they can be improved, could not be analyzed since the relation between $P(\bm{\nu})$ and $E(\bm{\nu})$ was not known~\cite{Heck13Blue}. 

There are two considerations for the error: 1) it should be low, 2) it should be as constant as possible, leading to white noise. The latter ensures that additional visual structures will not appear due to colored noise. We want to analyze how $U(\bm{\nu})$ and thus $P(\bm{\nu})$ should be shaped to achieve such a spectral profile for noise. For brevity, in the rest of the paper, we set $P(\bm{\nu}) = U(\bm{\nu}) + 1$, ignoring the Dirac delta at zero that does not contribute to $E(\bm{\nu})$.

\subsubsection*{Low energy for low frequencies} 
A fundamental property of anti-aliasing patterns such as blue noise patterns is that there should be a low energy low frequency region~\cite{Mitchell91Spectrally,Heck13Blue,Kailkhura16Stair}, i.e. $P(\bm{\nu})$ should be low and ideally zero for $\lVert \bm{\nu} \rVert < \nu_0$. This property is meant to limit the amount of noise $E(\bm{\nu})$ at lower frequencies. By expanding the convolution in Equation~\ref{eq:errorSpectrumAndP}, it can be easily shown for step blue-noise (Figure~\ref{fig:teaser}, left) where $P(\bm{\nu}) = U(\bm{\nu}) + 1 = 0$ for $\lVert \bm{\nu} \rVert < \nu_0$, and $1$ otherwise, that
\begin{equation}
\begin{aligned}
E_{STEP}(\bm{\nu}) 
&= \frac{1}{\lambda} \int_{-\infty}^{\infty} P_t(\bm{\nu} - \bm{\nu}') P_{STEP}(\bm{\nu}') d \bm{\nu}' \\
&= \frac{1}{\lambda} \int_{ \mathcal{D}_{\nu_0}^{\complement} } P_t(\bm{\nu} - \bm{\nu}') d \bm{\nu}' \\
&\leq \frac{1}{\lambda} \int_{-\infty}^{\infty} P_t(\bm{\nu} - \bm{\nu}') d \bm{\nu}' = E_{RND} (\bm{\nu}),
\label{eq:limitedNoiseWithBlueNoisePatterns}
\end{aligned}
\end{equation}
where $\mathcal{D}_{\nu_0}$ is the $d$-dimensional disk of radius $\nu_0$, $(\cdot)^{\complement}$ denotes the complement of a set, and $E_{STEP}$ and $E_{RND}$ are the errors when using step blue noise, and random sampling, respectively. In particular, for a band-limited function with $P_t(\bm{\nu}) = 0$ for $\lVert \bm{\nu} \rVert > \nu_0$, $E_{STEP} = 0$. However, in general, the difference between $E_{STEP}$ and $E_{RND}$ may not be very large especially when the function $P_t(\bm{\nu})$ has significant energy at higher frequencies. This can be seen in Figure~\ref{fig:CosSampling}, where we show examples of sampled images of a cosine wave $\cos ( 2 \pi \sqrt{\lambda} \nu_c y )$ of different frequencies $\nu_c$, and the corresponding $E(\bm{\nu})$, for different sampling patterns. For high frequencies such as $\nu_c = 1.5$, the low frequency region (implied by the reconstruction kernel $K$, marked with orange circles in the figure) of $E(\bm{\nu})$ for step blue noise contains as much energy as for random sampling, leading to similar levels of error in the sampled images.


In practice, having a large $\nu_0$ is still very important even when sampling non-band-limited functions, due to the stationarity of the point patterns considered (Section~\ref{sec:background}). As the patterns are translation invariant, each local patch of the function $t$ is sampled with a point distribution of the same characteristics. Hence, the same analysis can be carried out for each patch. The visual quality especially for smoother patches, where noise is visually very distractive, will thus be improved significantly by using a step-like profile. This is illustrated in Figure~\ref{fig:CosSampling} for $\nu_c = 0.35$, where stair blue noise and ds-wave sampling (a variation of our sampling patterns as we will discuss in Section~\ref{sec:optimizedPatterns}) with a higher $\nu_0$ than step blue noise and random sampling, result in much cleaner image content.


\subsubsection*{Limiting the maximum of $P$}
Another fundamental property utilized in the literature is that the maximum $m$ of $P$ should be limited~\cite{Heck13Blue,Kailkhura16Stair}. The intuition is that this will also limit the magnitude of and fluctuations in error. Indeed, we can easily show that 
\begin{equation}
\begin{aligned}
E(\bm{\nu})
&\leq \frac{m}{\lambda} \int_{-\infty}^{\infty} P_t(\bm{\nu} - \bm{\nu}') d \bm{\nu}'
= \frac{m}{\lambda} \int_{-\infty}^{\infty} P_t(\bm{\nu}) d \bm{\nu}.
\label{eq:limitingErrorWithM}
\end{aligned}
\end{equation}
Hence, normalized by the total energy of $T(\bm{\nu})$, the error in this case is bounded by $m/\lambda$ at \emph{every} frequency $\bm{\nu}$. 

This global maximum is also very important for limiting fluctuations in $E(\bm{\nu})$, i.e. avoiding colored noise. Example $E(\bm{\nu})$'s where such maxima add up to generate significant fluctuations in error for low frequencies are shown in Figure~\ref{fig:CosSampling}, stair blue noise sampling with $\nu_c = 0.75 - 0.85$. In this case, $P_t(\nu_1, \nu_2) = \alpha [ \delta ( \nu_2 - \sqrt{\lambda} \nu_c ) + \delta ( \nu_2 + \sqrt{\lambda} \nu_c ) ] \delta ( \nu_1 )$ for a constant $\alpha$, and hence the ratio of error to the total energy of the function $t$ is $E(\bm{\nu}) / \int_{-\infty}^{\infty} P_t(\bm{\nu}) d \bm{\nu} = \frac{1}{2\lambda} [ P ( \nu_1 - \sqrt{\lambda}\nu_c ) + P ( \nu_1 + \sqrt{\lambda}\nu_c ) ]$. In general for any function $t$, this ratio can fluctuate between $0$ and $m / \lambda$ at different frequencies, significantly disturbing the noise profile if the maximum $m$ is high. Such colored noise manifests itself as visually distinguishable secondary patterns in sampled images, as can be seen in the image reconstructions for stair blue noise with $\nu_c = 0.75 - 0.85$ in the figure, instead of white noise without a clear structure.

\subsubsection*{Minimizing local maxima of $P$}
Due to the constraints on $P(\bm{\nu})$ (Section~\ref{sec:background}), for point patterns with a larger $\nu_0$, $P(\bm{\nu})$ inevitably contains local maxima of decaying magnitude (as we will illustrate in Section~\ref{sec:optimizedPatterns}). Apart from limiting $m$, which determines the first maximum in $P(\bm{\nu})$, avoiding further local maxima is also beneficial, as these peaks can similarly sum up to cause further fluctuations in $E(\bm{\nu})$ due to the convolution in Equation~\ref{eq:errorSpectrumAndP}, albeit all smaller than $m$ as we illustrate in Figure~\ref{fig:CosSampling}. We will explore how we can shape the peaks such that we get an as small as possible global maximum $m$ and local maxima, while ensuring a certain $\nu_0$, by translating these into energies and constraints in a variational optimization based formulation for $P(\bm{\nu})$ in the next section.

\begin{figure}[t!]
  \centering
  \includegraphics[width=0.8\linewidth]{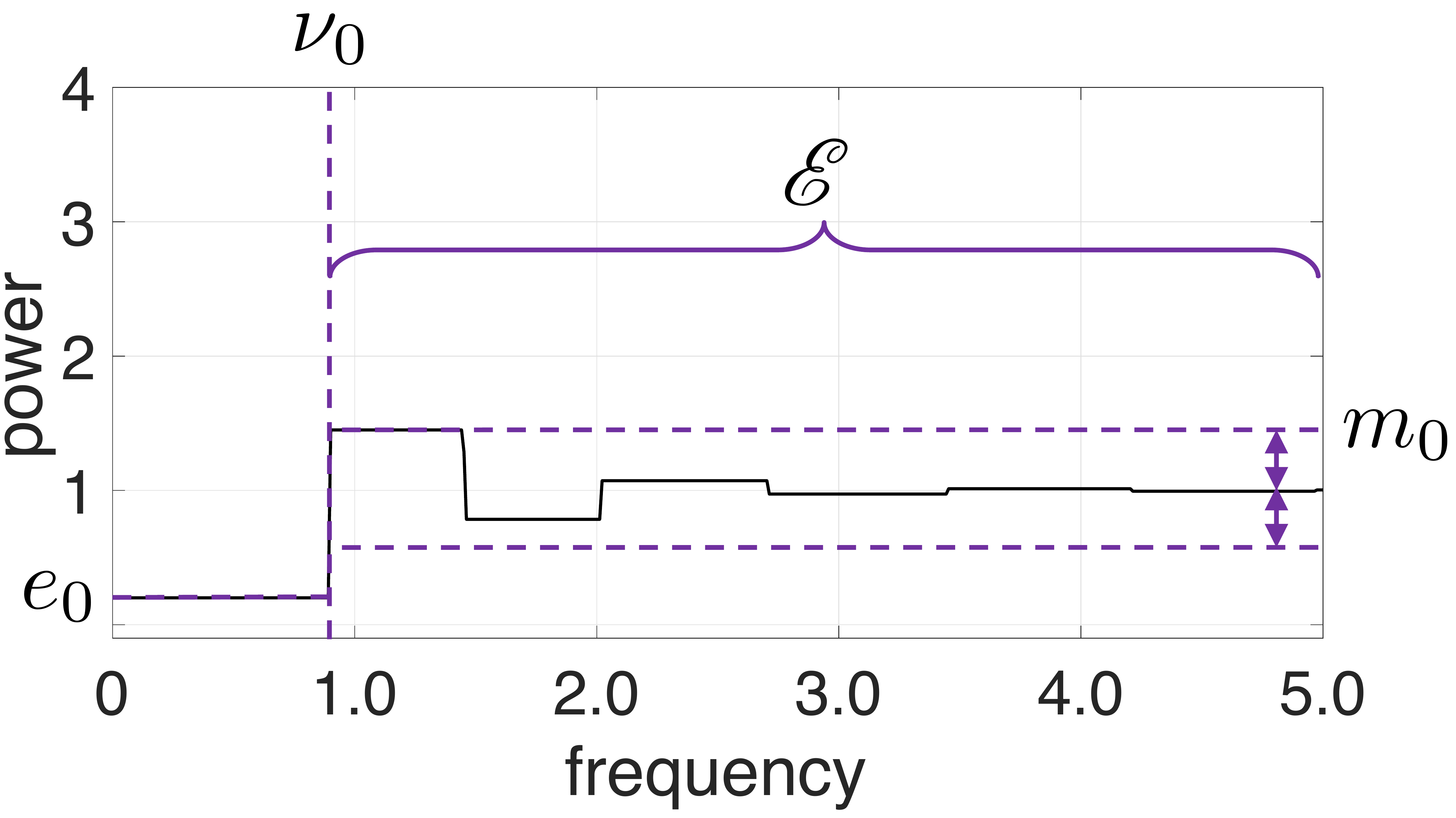}
 \caption{Realizable power spectra with anti-aliasing properties can be obtained by constrained variational optimization with constraints on the maximum value $e_0$ of the low frequency region $\nu < \nu_0$, deviation from $1$ for the high frequency region $\nu > \nu_0$ that implies the constraint $m_0$ on the maximum value of the spectrum, and an energy functional $\mathscr{E}$ that controls the shape of the high frequency region.
  \label{fig:EnergiesAndConstraints}}\vspace{-0mm}
\end{figure}

\section{Optimized Anti-aliasing Patterns}
\label{sec:optimizedPatterns}

The characteristics for $P$ as elaborated on in the last section can be imposed in addition to the realizability conditions (Equation~\ref{eq:PSRealizeConditions}), to obtain optimal sampling patterns with respect to these criteria. In this section, we formulate the associated variational optimization problem. This will allow us to synthesize optimal distributions with respect to the considered characteristics with numerical solution methods.

\begin{figure}[t!]
  \centering
  \includegraphics[width=0.99\linewidth]{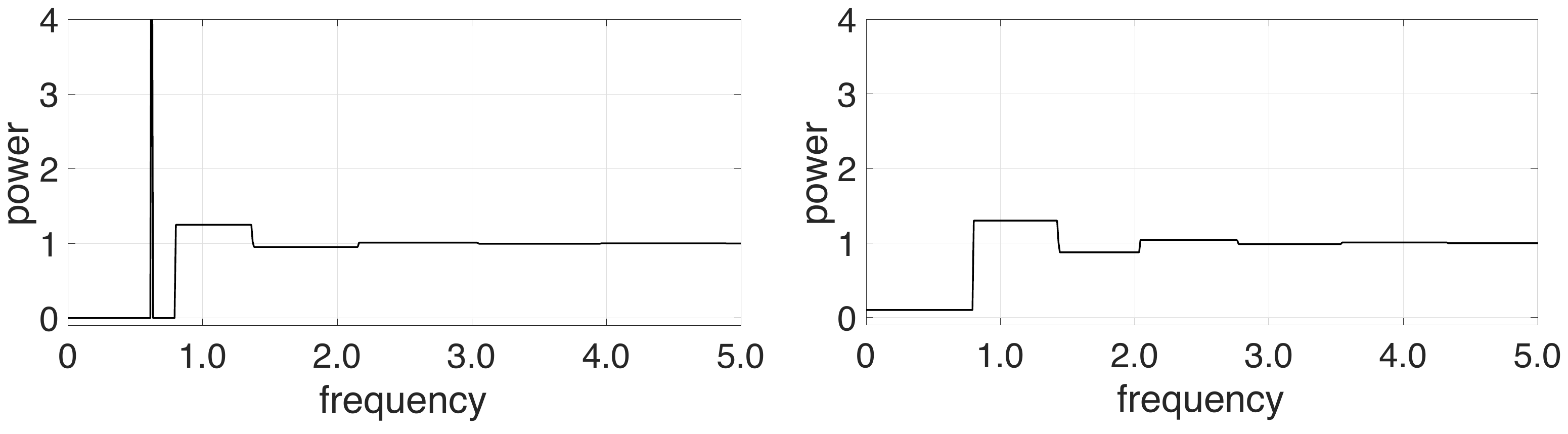}
 \caption{Power spectra generated with an integral-based low frequency constraint can lead to spikes for low frequencies (left). Directly limiting power for low frequencies ensures low aliasing (right).
  \label{fig:LowFreqConst}}\vspace{-0mm}
\end{figure}

\subsection{Sampling as Constrained Variational Optimization}

We start by making the effect of density $\lambda$ on the problem explicit, and factor it out from the optimization. This can be achieved by working with normalized spatial and frequency coordinates. We start by defining $f(\mathbf{r}) = u (\mathbf{r} / \lambda^{1/d} ) / \lambda$. By the scaling property of Fourier transform, we can write $F(\bm{\nu}) = U(\bm{\nu} \lambda^{1/d})$. Substituting these into the expressions for $g$ and $P$ (Equation~\ref{eq:powerSpectrumAndPCFInTermsOfU}), we get $g(\mathbf{r} / \lambda^{1/d}) = f(\mathbf{r}) + 1$, and $P(\bm{\nu} \lambda^{1/d}) = F(\bm{\nu}) + 1$ (ignoring $\delta(\bm{\nu})$ as before, as it does not contribute to $E(\bm{\nu})$). Then, in normalized coordinates, we can write the conditions $g(\mathbf{r} / \lambda^{1/d}) \geq 0$ and $P(\bm{\nu} \lambda^{1/d}) \geq 0$ as
\begin{equation}
\begin{aligned}
f(\mathbf{r}) + 1 \geq 0, \hspace{5mm} F(\bm{\nu}) + 1 \geq 0.
\label{eq:PSRealizeConditionsInTermsOfF}
\end{aligned}
\end{equation}
Thus, the constraints become independent of the intensity $\lambda$ of the point process. We will work with $g$ and $P$ in normalized coordinates unless stated otherwise, and set $g(\mathbf{r}) = f(\mathbf{r}) + 1$, and $P(\bm{\nu}) = F(\bm{\nu}) + 1$. Absolute spatial coordinates are thus given by multiplying the reported $\mathbf{r}$ with $1 / \lambda^{1/d}$, and absolute frequencies by multiplying the reported $\bm{\nu}$ with $\lambda^{1/d}$.


\begin{figure*}[t!]
  \centering
  \includegraphics[width=0.99\linewidth]{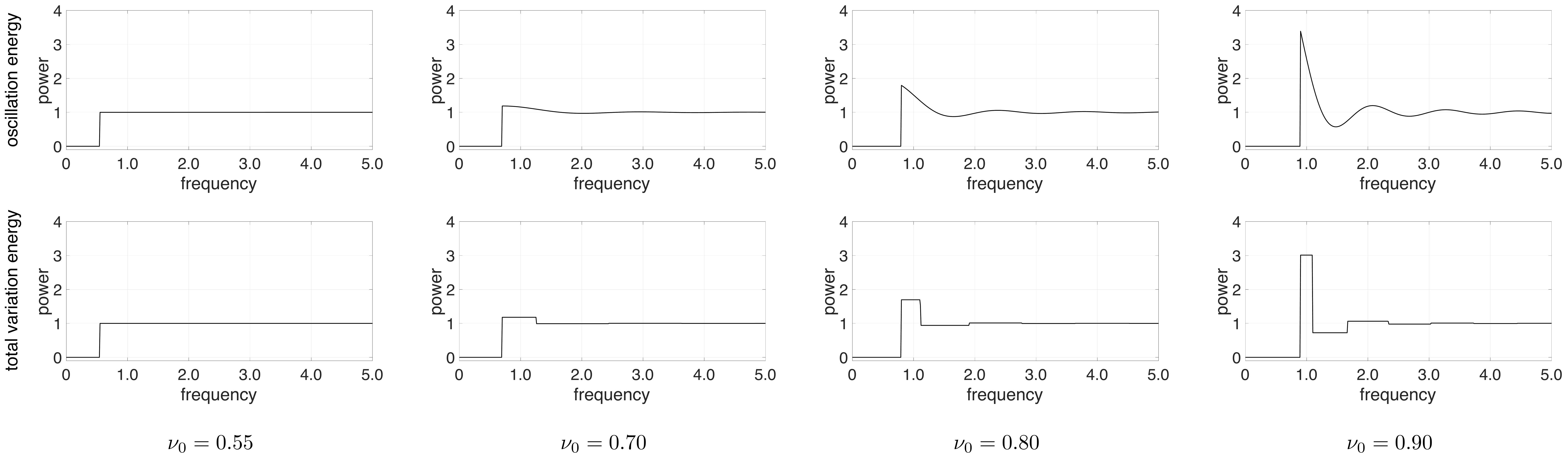}
 \caption{Power spectra generated by minimizing oscillation (top) and total variation (bottom) energies under the low frequency constraint (Section~\ref{sec:constraintsAndEnergies}). Minimizing total variation provides a non-traditional spectrum with a decaying square wave form and lower maximum values for the power spectra.
  \label{fig:OscillationVsTV}}\vspace{-0mm}
\end{figure*}

Although the complete analysis in the rest of the paper can be carried out for stationary point processes in $\mathbf{R}^d$, we will consider the important case of image sampling with non-adaptive anti-aliasing distributions as in previous works~\cite{Dippe85Antialiasing,Heck13Blue,Kailkhura16Stair}. This implies that the point processes considered are isotropic, generating rotation and translation invariant distributions. In this case, $g$ and thus $P$ is radially symmetric such that $g(\mathbf{r}) = g(\Vert \mathbf{r} \Vert) = g(r)$, $P(\bm{\nu}) = P(\Vert \bm{\nu} \Vert) = P(\nu)$, and all Fourier transforms in the definitions above turn into Hankel transforms $\mathscr{H}$. In particular, we have $F = \mathscr{H}[f]$, or equivalently $f = \mathscr{H}[F]$. The Hankel transform is defined for any dimensions. For our case of sampling the image plane, we use the Hankel transform for $d = 2$ dimensions.

Hence, the problem becomes finding a 1D function $F$ with the above non-negativity constraints in Equation~\ref{eq:PSRealizeConditionsInTermsOfF}, and additional properties we impose. These properties can either be set as hard inequality constraints $\mathscr{C}[F(\nu)] \geq 0$, or energies that we minimize for. We can thus formulate the following constrained variational minimization problem on $F$, to find a realizable and desirable $P$
\begin{equation}
\begin{gathered}
\min \mathscr{E}[F] \\ 
F + 1 \geq 0, \hspace{5mm} \mathscr{H}[F] + 1 \geq 0, \hspace{5mm} \mathscr{C}[F] \geq 0.
\end{gathered}
\label{eq:continuousOptimizationProblem}
\end{equation}

\subsection{Constraints and Energies for Anti-aliasing}
\label{sec:constraintsAndEnergies}

By changing the energy functional $\mathscr{E}$ and constraints $\mathscr{C}$,  we can tune the properties of $F(\nu)$ and thus $P(\nu) = F(\nu) + 1$.  We summarize the properties shaped by the constraints and energies considered in Figure~\ref{fig:EnergiesAndConstraints}. These closely follow the criteria analyzed in Section~\ref{sec:theoreticalAnalysisOfError}, with a low energy low frequency region, bounded global maximum and local maxima.


\subsubsection*{Low Frequency Constraint}

We start with the property that $P(\nu)$ has small energy for $\nu < \nu_0$ for a given $\nu_0$ (we call this region as the \emph{low frequency region}, and $\nu > \nu_0$ as the \emph{high frequency region}). This can be imposed with a direct constraint of the form $\frac{1}{\nu_0} \int_{0}^{\nu_0} P(\nu) d\nu \leq e_0$, for a limit $e_0$. Similar terms have been used to quantify the energy in the low frequency region in previous works~\cite{Heck13Blue,Kailkhura16Stair}. However, this does not limit the value of $P(\nu)$ at a given frequency, and hence $P(\nu)$ can grow very large, leading to severely high error at certain low frequencies, and hence significant fluctuations in the spectrum. An example spectrum generated with this integral constraint is shown in Figure~\ref{fig:LowFreqConst}, left. Instead, we propose to directly limit the spectrum for the low frequency region with
\begin{equation}
\begin{aligned}
F(\nu) + 1 \leq e_0 \hspace{5mm} \nu < \nu_0.
\label{eq:v0Constraint}
\end{aligned}
\end{equation}
This ensures that error will be bounded at all low frequencies (Figure~\ref{fig:LowFreqConst}, right). Ideally, $e_0 = 0$, and thus $P(\nu) = 0$ for $\nu < \nu_0$. We will see in the next section that $e_0$ and hence noise for low frequencies can be traded off with aliasing at higher frequencies. For the analysis in this section, we assume $e_0 = 0$.

\subsubsection*{Oscillation Energy}

Another desired property of $P(\nu)$ is that it should not have high global and local maxima. This can be imposed in several ways. Previous works~\cite{Heck13Blue} have considered measuring squared deviation of $P(\nu)$ from $1$, which can be written as the following energy
\begin{equation}
\begin{aligned}
\mathscr{E}[F]  = \int_{\nu_0}^{\infty} F^2(\nu) d\nu.
\label{eq:oscillationEnergy}
\end{aligned}
\end{equation}
Minimizing this energy with the realizability constraints and the low frequency constraint above for different $\nu_0$'s, we get the spectra in Figure~\ref{fig:OscillationVsTV}, top. We get a perfectly zero region for $\nu < \nu_0$, and peaks of decaying magnitude for higher frequencies. This is a typically encountered profile for blue noise patterns, except for two recent works~\cite{Heck13Blue,Kailkhura16Stair}. For $\nu_0 < \sqrt{1 / \pi}$, which is the theoretical limit for a step-noise profile, we get a perfect step shape. Larger $\nu_0$ leads to oscillations, with the magnitude of the first peak determining the global maximum of $P$. As $\nu_0$ is increased, the maximum also gets larger.


\subsubsection*{Total Variation Energy}

For $\nu > \nu_0$ and $\nu_0 > \sqrt{1/\pi}$, $P$ will inevitably deviate from $1$ with one or more peaks, before it (possibly) converges to $1$~\cite{Heck13Blue}. An alternative way of limiting the magnitudes of these peaks is to minimize total variation energy. This can be visualized as minimizing the length of the path traveled by a point when projected onto the $P$-axis, as it moves along the curve $P(\nu)$ from $\nu_0$ to $\infty$. The resulting energy is given by
\begin{equation}
\begin{aligned}
\mathscr{E}[F]  = \int_{\nu_0}^{\infty} | F'(\nu) | d\nu.
\label{eq:totalVarEnergy}
\end{aligned}
\end{equation}
We show power spectra generated by minimizing this energy under the low frequency constraint and realizability conditions in Figure~\ref{fig:OscillationVsTV}, bottom. The spectra now mostly contain raised rectangular regions instead of peaks, i.e. a decaying square wave. This is due to the sparse gradients introduced by total variation. Heights of rectangular regions, and thus the maxima of $P$ are smaller than when minimizing the oscillation energy above.

\begin{figure}[t!]
  \centering
  \includegraphics[width=0.99\linewidth]{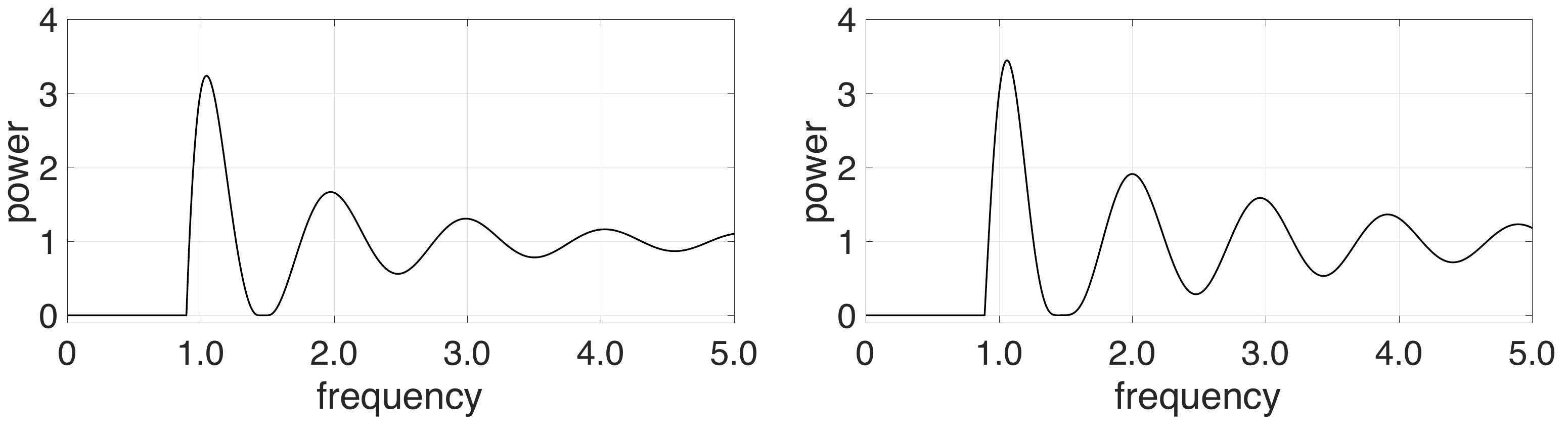}
 \caption{Power spectra generated by minimizing smoothness energies (left: Dirichlet, right: Laplacian energy) for $\nu_0 = 0.9$. We get higher peaks and thus worse characteristics than other energies considered.
  \label{fig:SmoothnessEnergy}}\vspace{-0mm}
\end{figure}

\subsubsection*{Smoothness Energy}

We further experimented with smoothness energies, the Dirichlet energy $\mathscr{E}[F]  = \int_{\nu_0}^{\infty} | F'(\nu) |^2 d\nu$, and Laplacian energy $\mathscr{E}[F]  = \int_{\nu_0}^{\infty} | F''(\nu) |^2 d\nu$. We show the resulting spectra in Figure~\ref{fig:SmoothnessEnergy}. The spectra in these cases are worse with higher peaks than with oscillation or total variation energy.

\subsubsection*{Maximum Constraint}

\begin{figure}[t!]
  \centering
  \includegraphics[width=0.99\linewidth]{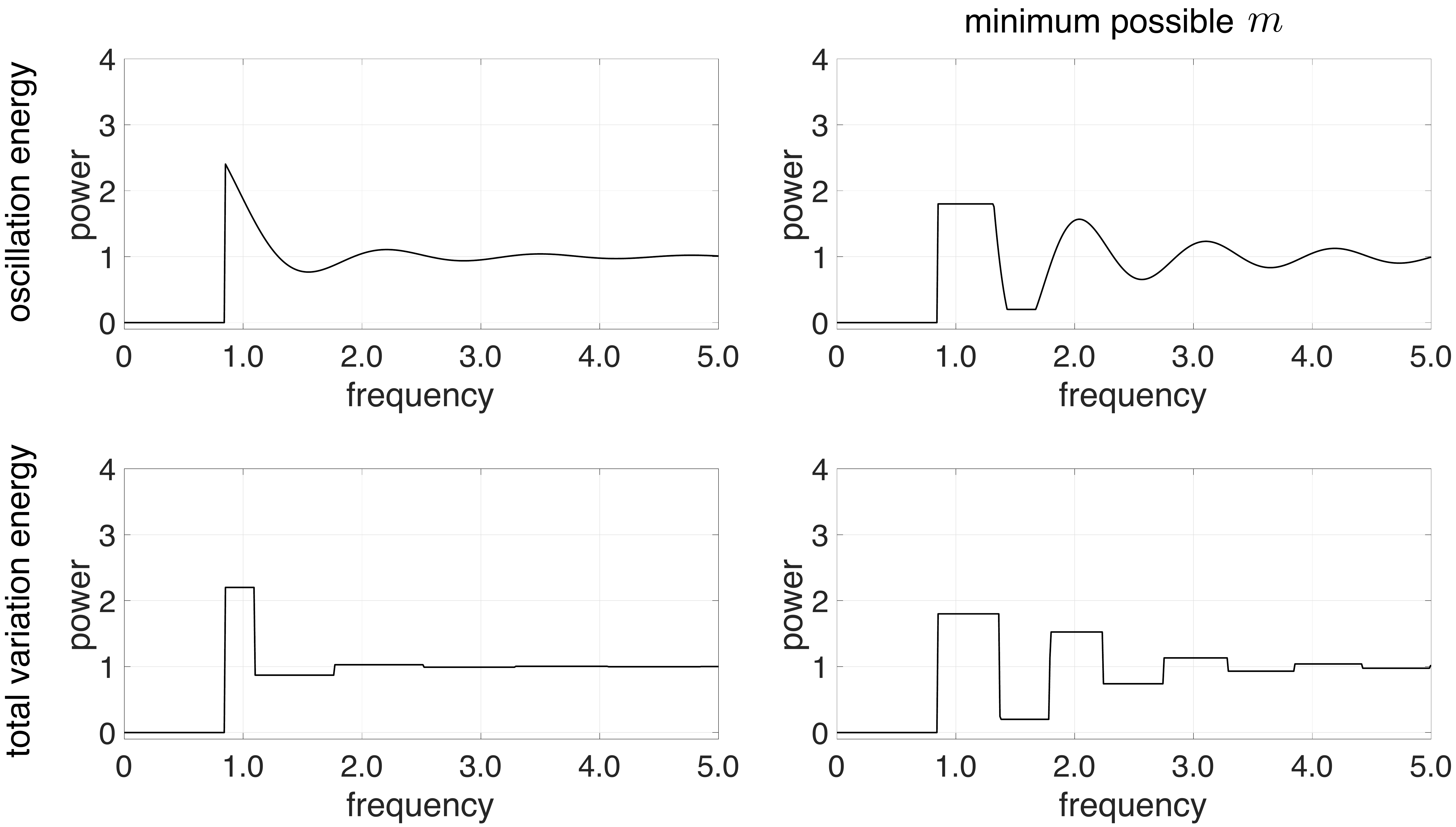}
 \caption{Power spectra generated by minimizing oscillation and total variation energies without (left) and with (right) the maximum constraint (Equation~\ref{eq:maxConstraint}) for the optimal (minimum possible) $m_0$ and thus $m$ ($\nu_0 = 0.85$ for all cases). 
  \label{fig:OscillationVsTVMaxConst}}\vspace{-0mm}
\end{figure}

Although total variation energy leads to peaks of smaller magnitude, it might still be possible to further reduce the global maximum $m$ of $P$. The same is true for all energies. Having a small $m$ is an important factor to avoid colored noise and hence aliasing, as elaborated on in Section~\ref{sec:analysisOfAntialiasingProperties}. To achieve a smaller $m$, we limit the magnitude of deviations from $1$ with the following constraint
\begin{equation}
\begin{aligned}
|  P(\nu) - 1 | = |  F(\nu) |  \leq m_0 - 1 \hspace{5mm} \nu > \nu_0.
\label{eq:maxConstraint}
\end{aligned}
\end{equation}
In practice, this constraint is equivalent to $P(\nu) \leq m_0$, as for all power spectra in previous works and in this work, the first peak has the largest  $|  P(\nu) - 1 |$, and at that peak $P(\nu) > 1$ (e.g. Figure~\ref{fig:OscillationVsTV}). Of course, not all $\nu_0$ - $m_0$ combinations are realizable (we will elaborate more on this point in the next section). In order to define the range of possible $m_0$ for a given $\nu_0$, we can find the minimum possible $m_0$ by an exhaustive search. We show this constraint imposed on the spectra for oscillation energy and total variation energy in Figure~\ref{fig:OscillationVsTVMaxConst}. For both energies, reducing $m_0$ comes at the cost of a larger number of oscillations, albeit all with smaller magnitudes, hence not leading to significant aliasing. 


%




\subsection{Optimized Sampling Patterns}

The analysis above suggests that total variation energy leads to better profiles for power spectra, with lower global and local maxima. Without the maximum constraint (Equation~\ref{eq:maxConstraint}), the maximum with oscillation energy is larger (Figure~\ref{fig:OscillationVsTVMaxConst}, left), while imposing the maximum constraint results in further peaks of higher magnitudes than those with total variation energy as in Figure~\ref{fig:OscillationVsTVMaxConst}, right (please see the supplementary material for more spectra with total variation and oscillation energies). Due to the flatter shape of the spectrum, noise is introduced for a larger range of unrepresentable high frequencies with total variation energy. However, such incoherent noise is preferred to colored noise caused by higher maxima in power spectra with oscillation energy. Hence, we focus on total variation (Equation~\ref{eq:totalVarEnergy}) as the energy in this paper. Due to its shape resembling a decaying square wave, we call the resulting pattern as \emph{ds-wave} sampling.

\begin{figure*}[t!]
  \centering
  \includegraphics[width=0.99\linewidth]{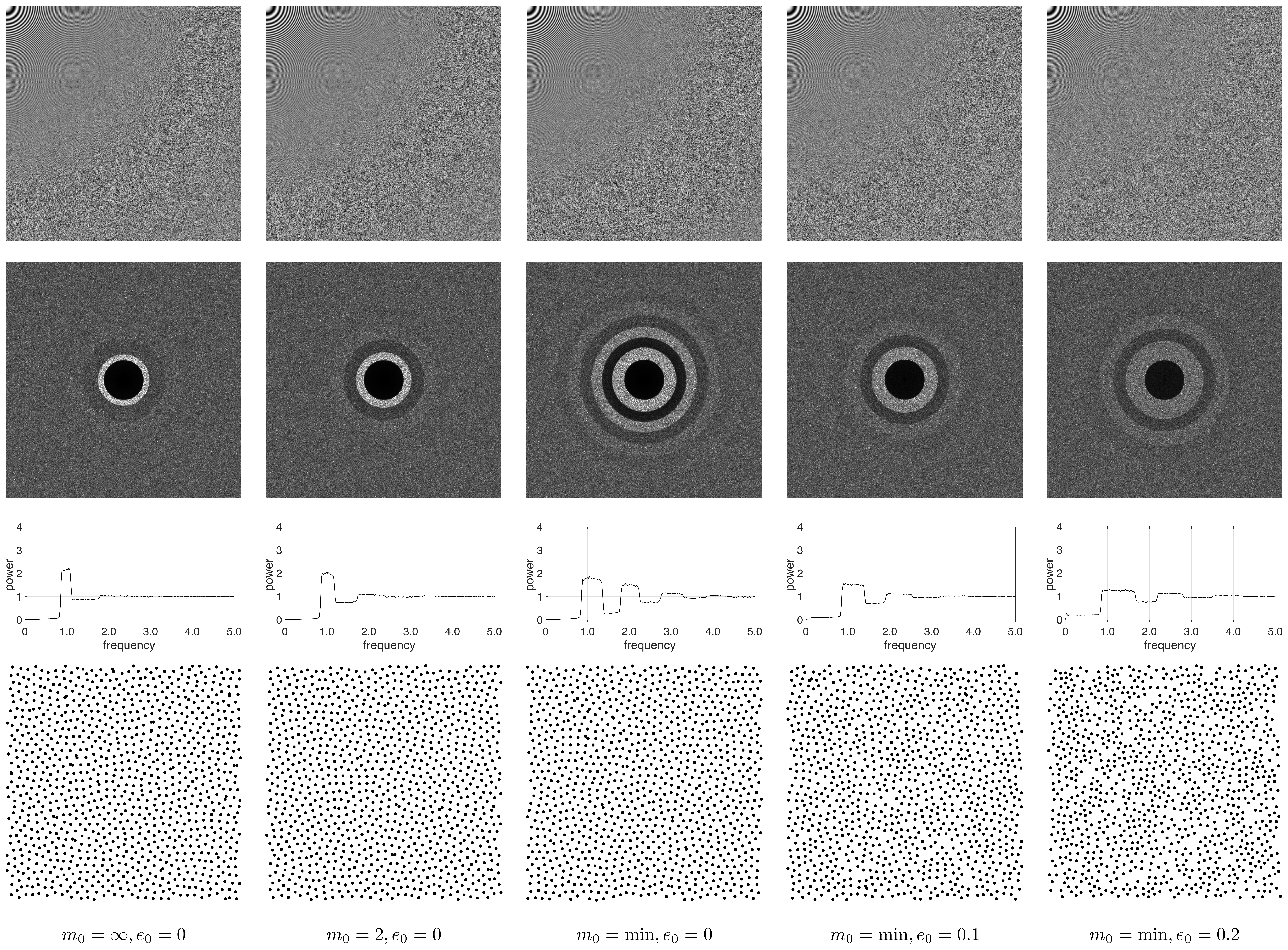}
 \caption{Optimized power spectra with total variation energy and $\nu_0 = 0.85$, for different $m_0$ and $e_0$ values. From top to bottom: zone plate image sampled with the generated point distributions, estimated 2D power spectrum, 1D power spectrum, an example distribution.
  \label{fig:OptimumPatterns}}\vspace{-0mm}
\end{figure*}

The use of the maximum constraint depends on the gain we obtain, i.e. how much lower the maximum of the power spectrum is with this constraint. We show estimated power spectra without ($m_0 = \infty$) and with the maximum constraint for $m_0 = 2$, and $m_0 = \mbox{min}$ (the minimum possible $m_0$) in Figure~\ref{fig:OptimumPatterns}. These are computed as the empirical power spectra of generated point distributions (we elaborate more on this in the next section). As we use total variation energy, we already get low maxima, hence see only a marginal improvement. In general, it is possible to tune $m_0$ depending on how critical this improvement is for the application, but this requires a search among different $m_0$ values, and some additional local maxima appear in the spectrum.

\begin{figure}[t!]
  \centering
  \includegraphics[width=0.99\linewidth]{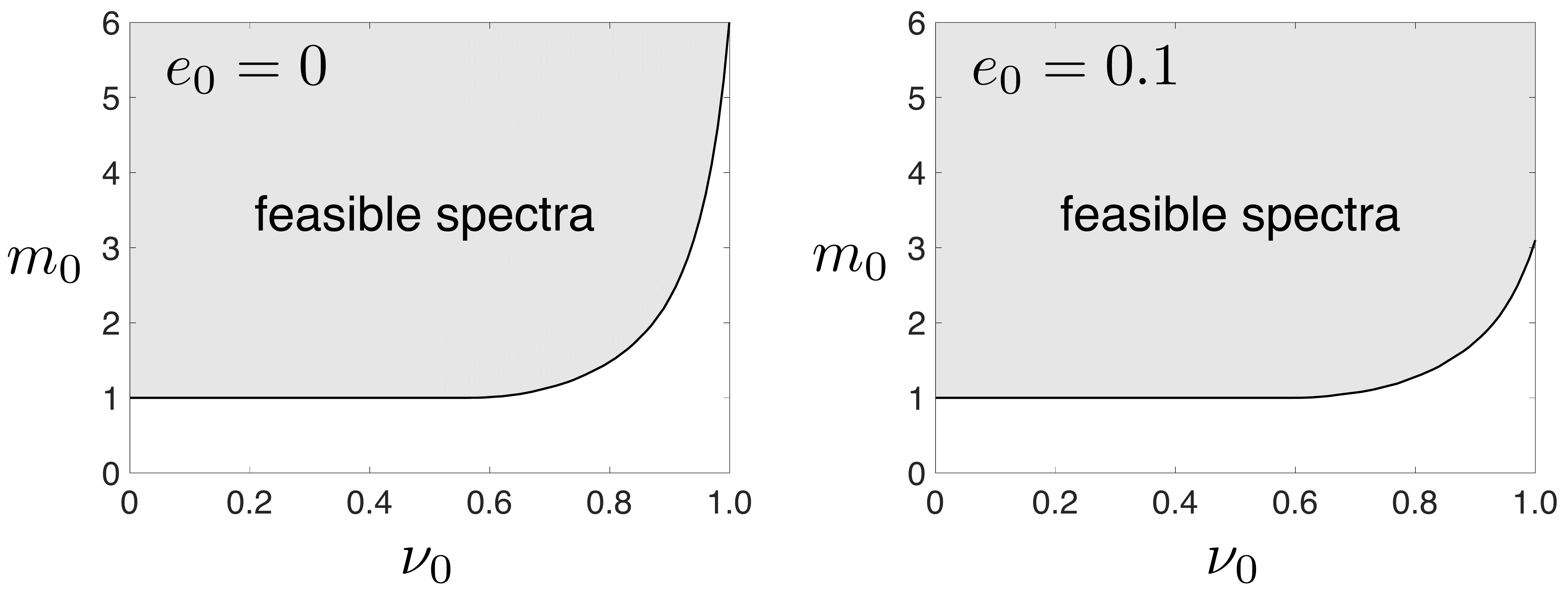}
 \caption{Feasible regions for realizable power spectra in the $\nu_0$ - $m_0$ space for different $e_0$'s. No patterns can have power spectra outside this region.
  \label{fig:OptimumV0M0Graph}}\vspace{-0mm}
\end{figure}

Recent works~\cite{Heck13Blue,Kailkhura16Stair} explore minimizing maximum of power spectra for a given $\nu_0$ and $e_0$. However, as they use pre-defined parametric families of functions, it is not possible to achieve the minimum possible maximum. By utilizing the proposed optimization framework, we can derive the \emph{minimal} maximum (up to numerical accuracy). For fixed $\nu_0$ and $e_0$, we try to optimize any of the above energies for various $m_0$'s, and take the minimum that leads to a feasible solution. The resulting space of feasible $\nu_0$ - $m_0$ pairs are shown in Figure~\ref{fig:OptimumV0M0Graph} for $e_0 = 0$ and $e_0 = 0.1$. Note that these results are general and independent of the form of the power spectrum. The minimum possible $m_0$ stays at $1$ for $\nu < \sqrt{1/\pi}$ as expected, and becomes increasingly more sensitive to $\nu_0$ for large $\nu_0$ values. It is not possible to go beyond $\nu_0 = 1$, as this is the $\nu_0$ of regular sampling. We also observe this in practice when we compute the feasible region. We can use the feasible region as a benchmark for how patterns perform for anti-aliasing. 


As Figure~\ref{fig:OptimumV0M0Graph}, right, shows, increasing low frequency noise with $e_0 = 0.1$ significantly reduces $m_0$, especially for high values of $\nu_0$. We show power spectra obtained for $e_0 = 0.1$ and $e_0 = 0.2$ in Figure~\ref{fig:OptimumPatterns}. Although the constraint is $P(\nu) \leq e_0$ for $\nu < \nu_0$, the optimizations result in $P(\nu) = e_0$ for all $\nu < \nu_0$. For higher $e_0$'s, power spectra get flatter, and thus zone plate images show reduced aliasing artifacts for high frequencies. This comes at the cost of higher levels of noise introduced into low frequencies, as visible in low frequency parts of the zone plate images. The generated point distributions reveal the source of this low frequency noise: they become increasingly more random for larger values of $e_0$.




\subsection{Implementation}
\label{sec:implementation}

We discretize the problem in Equation~\ref{eq:continuousOptimizationProblem} with standard techniques from numerical analysis. As the function $F$ to optimize for is $1$-dimensional, a simple discretization with regular sampling is used. The derivative operators are then discretized with finite differences, and integrals in the energies are approximated with the trapezoidal rule. Hankel transform is discretized with an accurate approximation based on the trapezoidal rule (e.g. \cite{Cree93Algorithms}, Equation 6).

We experimented with various ranges and sampling rates for $F$ and $\mathscr{H}[F]$. In all of our experiments, a sample spacing of $0.01$ was sufficient for accurate numerical results. In order to test the accuracy of the resulting spectra with this spacing, we take $100$ spectra with randomly chosen parameters ($\nu_0 \in [0.5, 1)$, and $e_0 \in [0, 0.2]$ to avoid having many step-like spectra). We then compute the average root mean squared difference between each spectrum for a sample spacing of $0.01$ and $0.001$, resulting in $6.1 \hspace{1mm} 10^{-3}$ average difference. For $P$, we sample the range $\nu \in [0, 10]$, since well before $\nu = 10$, $F$ converges to $0$ (and hence $P = F + 1$ converges to $1$) for all patterns in our experiments. The average absolute deviation of $F$ from $0$ for $\nu \in [9, 10]$ for $100$ random spectra is $7.74 \hspace{1mm} 10^{-5}$. The same, however, is not true for $\mathscr{H}[F]$, which determines the PCF $g$. For $\mathscr{H}[F]$, we thus sample almost the full range of possible distances for the unit toroidal domain we consider, $[0, 0.5]$ in absolute coordinates. Note, however, that in practice this is not strictly needed as we only require $g(r) = \mathscr{H}[F](r) + 1 \geq 0$, and  $g(r)$ does not oscillate significantly beyond a limited range of $r$'s.



Hankel transform is a linear operator and hence all constraints, including the realizability conditions (Equation~\ref{eq:PSRealizeConditionsInTermsOfF}), are linear inequality constraints. The oscillation (Equation~\ref{eq:oscillationEnergy}) and smoothness energies turn into quadratic forms when discretized. For these energies, the discrete problem thus becomes quadratic programming. Minimization with total variation energy (Equation~\ref{eq:totalVarEnergy}) can be formulated as linear programming, using well-known results from optimization. Both problems are thus convex and easy to solve with any modern optimization package. We use built-in Matlab functions for optimization.

There are several techniques for synthesis of point patterns based on PCF or power spectrum~\cite{Zhou12Point,Oztireli12Analysis,Heck13Blue,Wachtel14Fast,Ahmed15AA,Kailkhura16Stair}. We experimented with several of these approaches that focus on accuracy~\cite{Oztireli12Analysis,Heck13Blue,Kailkhura16Stair}, and got similar results. We hence use the PCF-based matching technique of Heck et al.~\shortcite{Heck13Blue} for all experiments in the paper and the supplementary material, due to the efficient implementation available. We use the default parameters for that algorithm, and the same discretization for PCF as we describe above.


%% file: TexFiles/Results.tex
\section{Evaluation and Analysis}

To evaluate the performance of our sampling patterns in practice, we analyze power spectra, and illustrate their anti-aliasing properties on sampled images (for more results, please see the supplementary material). 

\subsubsection*{Evaluation}

For estimating the power spectrum of a pattern, we generate $10$ point distributions with a matching spectrum, each with $4096$ points. The empirical power spectra of these distributions, computed with Equation~\ref{eq:powerSpectrumNdDefinition} and their radial averages, are averaged to generate all estimated 2D and 1D power spectra in this paper and the supplementary material. For other sampling patterns that start from a theoretical power spectrum (step blue noise, single-peak blue noise~\cite{Heck13Blue}, and stair blue noise~\cite{Kailkhura16Stair}), we use the same point distribution synthesis algorithm~\cite{Heck13Blue} with the same settings as in Section~\ref{sec:implementation}.

Unless stated otherwise, we use distributions with $16384$ ($128 \times 128$) points at $1$ sample per pixel (spp) for all test images except the zone plate images. We use the zone plate function ($[1 + \cos(\alpha \Vert \mathbf{x} \Vert^2)]/2$, $\mathbf{x} \in \mathcal{V}$) as a benchmark test image, since it reveals aliasing at different frequencies without the masking effect due to local structures~\cite{Heck13Blue,Kailkhura16Stair}. These images are sampled at $2$ spp (please see the supplementary material for zone plate images with $1, 2, 4$ spp). All images are reconstructed with a low-pass filter (we use the $3\times3$ Gaussian filter) and resampled to a regular grid. This filter and others used in the literature retain some of the frequencies that cannot be represented with the pixel grid, but ensure a clear visualization of noise. In practice, we only noticed faint and distinct secondary rings for zone plate images when using more than $1$ samples per pixel due to this filtering, as visible e.g. in Figure~\ref{fig:teaser}.


\begin{figure}[t!]
  \centering
  \includegraphics[width=0.99\linewidth]{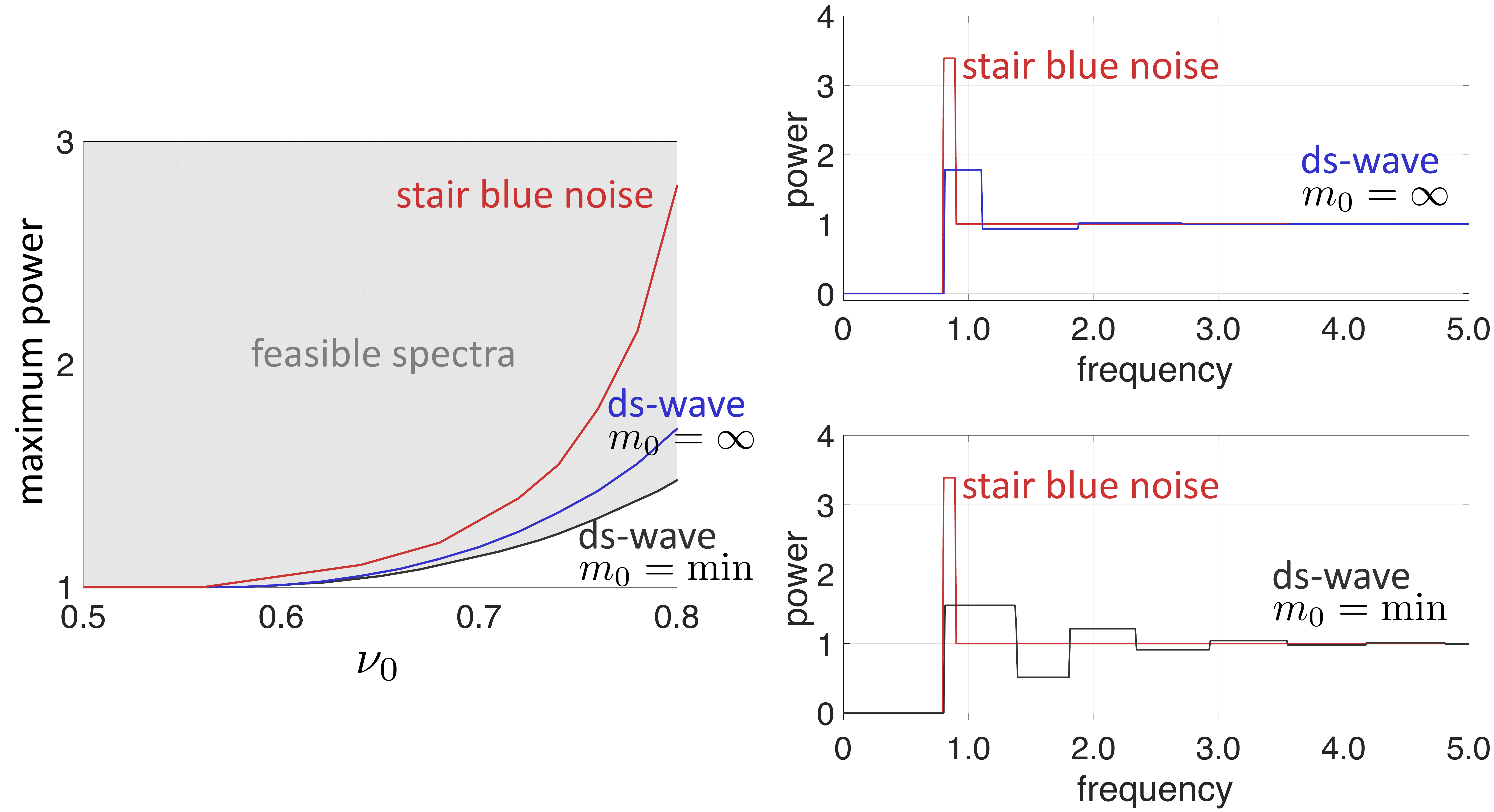}
 \caption{(Left) For each $\nu_0$, we show the maximum value of the power spectrum for ds-wave and stair blue noise~\cite{Kailkhura16Stair} sampling. (Right) Comparisons of spectra of stair blue noise and ds-wave sampling at the maximum achievable $\nu_0 = 0.81$ for stair blue noise.
  \label{fig:ComparisonsStair}}\vspace{-0mm}
\end{figure}

\begin{figure*}[t!]
  \centering
  \includegraphics[width=0.99\linewidth]{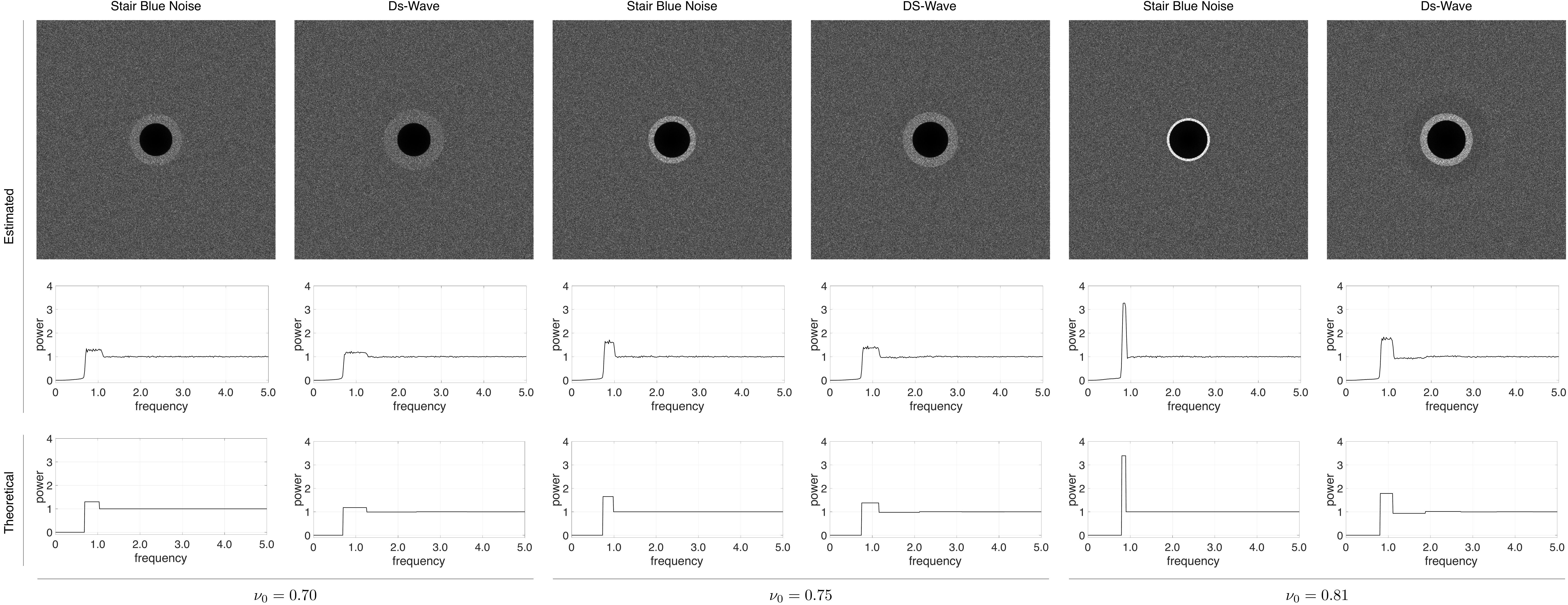}
 \caption{Estimated and theoretical power spectra for stair blue noise~\cite{Kailkhura16Stair} and ds-wave sampling for different $\nu_0$ values ($e_0 = 0$, $m_0 = \infty$). For all cases, ds-wave sampling results in a lower maximum for the power spectra.
  \label{fig:ComparisonsStairSynthesized}}\vspace{-0mm}
\end{figure*}

\subsubsection*{Ds-wave and Stair Blue Noise} 

We analyze the properties of our ds-wave sampling and compare them to those of the state-of-the-art stair blue noise sampling~\cite{Kailkhura16Stair} in Figure~\ref{fig:ComparisonsStair}. For stair blue noise, at each $\nu_0$, we find the minimum value for the maximum $m$ of the power spectrum with an exhaustive search over the parameters, as described by Kailkhura et al.~\shortcite{Kailkhura16Stair}. Since stair blue noise has zero energy for the low frequency region, we set $e_0 = 0$ for all comparisons.

Figure~\ref{fig:ComparisonsStair}, left shows that for all $\nu_0 > \sqrt{1/\pi}$ (the theoretical limit for step blue noise), ds-wave sampling has a lower $m$ than stair blue noise, with the difference getting larger for larger $\nu_0$. Ds-wave sampling with $m_0 = \mbox{min}$ achieves the optimal $m$, by definition. But even without the maximum constraint (i.e. $m_0 = \infty$), ds-wave sampling has $m$ very close to the optimum. Note that the maximum $\nu_0$ we could obtain for stair blue noise sampling for $m \leq 100$ is $0.81$, and thus we show the range $\nu_0 \in [0.5, 0.8]$. In Figure~\ref{fig:ComparisonsStair}, right, we plot the spectra for $\nu_0 = 0.81$, illustrating the significant difference between ds-wave and stair blue noise for both $m_0 = \infty$ and $m_0 = \mbox{min}$. We show further theoretical, and estimated 2D and 1D spectra for lower $\nu_0$'s in Figure~\ref{fig:ComparisonsStairSynthesized} ($m_0 = \infty$). For all cases, the theoretical spectra of ds-wave sampling can be realized reliably, with a lower maximum than stair blue noise, and almost no further oscillations. The difference is especially significant for higher values of $\nu_0$. Please see the supplementary material for more examples of theoretical and estimated spectra.

\begin{figure*}[t!]
  \centering
  \includegraphics[width=0.80\linewidth]{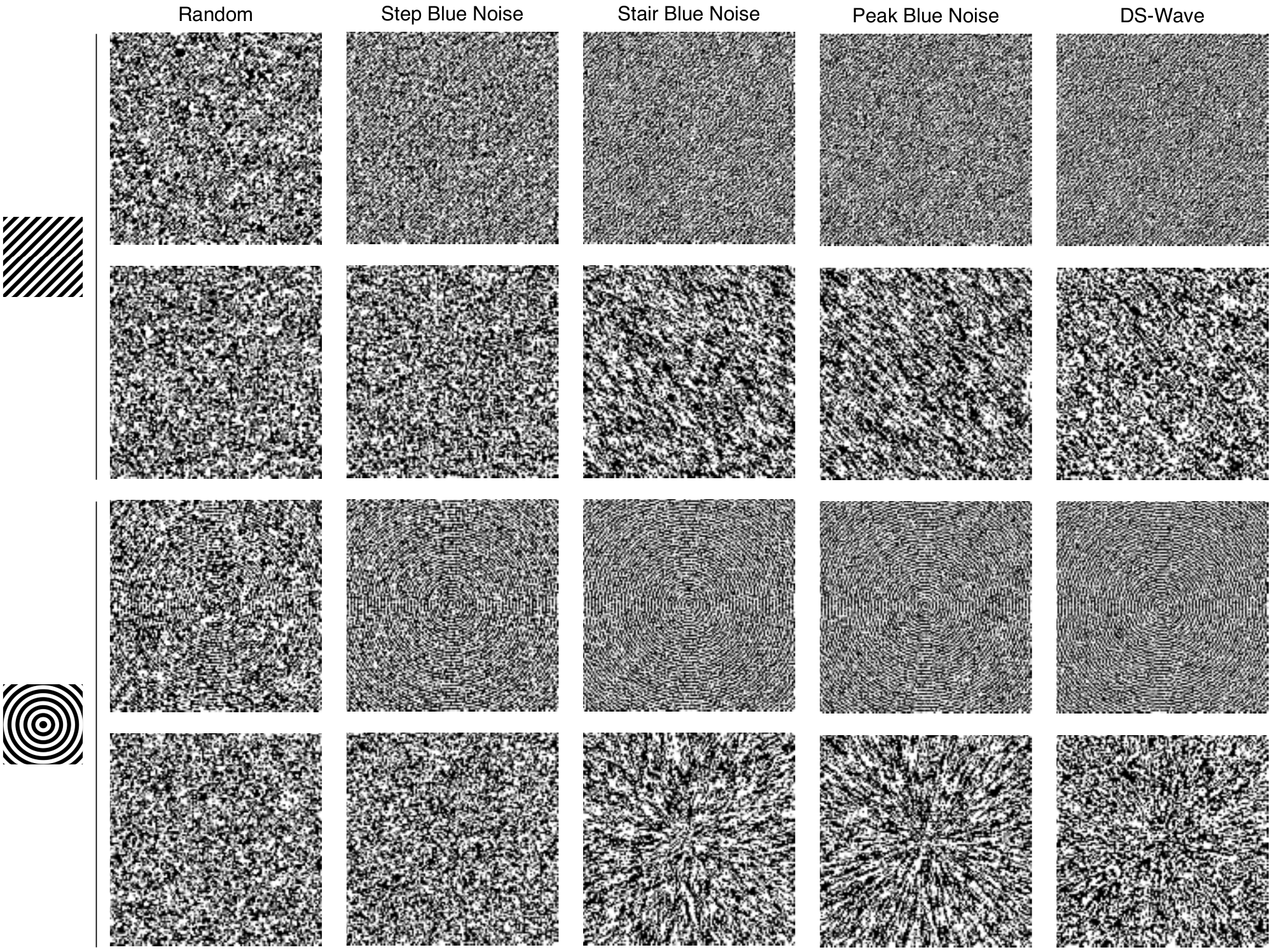}
 \caption{State-of-the-art sampling patterns such as stair blue noise~\cite{Kailkhura16Stair} and peak blue noise~\cite{Heck13Blue} result in less noise than random and step blue noise sampling for repeated structures with a low frequency (top rows), at the cost of introducing colored noise and hence secondary patterns when sampling images with repeated structures of higher frequencies (bottom rows). Ds-wave sampling leads to cleaner reconstructions, and smoothly degrades to as white as possible noise for repeated structures with high frequencies, avoiding visible secondary patterns. (Visualizations of the original stripe patterns in sampled images are shown on the left. The patterns repeat every $1/\nu_c$ pixels, with $\nu_c = 0.5$ (top rows) and $\nu_c = 0.9$ (bottom rows). Peak blue noise is with an effective $\nu_0 = 0.85$ as explained in the text, stair blue noise is with $\nu_0 = 0.81$ (maximum possible), and ds-wave is with $\nu_0 = 0.85$, $m_0 = \mbox{min}$.)
  \label{fig:ComparisonPatterns}}\vspace{-0mm}
\end{figure*}

The practical utility of these results is illustrated in Figures~\ref{fig:teaser} ($\nu_0 = 0.80$,  $m_0 = \infty$), ~\ref{fig:CosSampling} ($\nu_0 = 0.80$,  $m_0 = \infty$), and~\ref{fig:ComparisonPatterns} ($\nu_0 = 0.85$,  $m_0 = \mbox{min}$). In Figure~\ref{fig:teaser}, the zone plate image reveals frequencies that are mapped to colored noise and hence secondary patterns due to the high magnitude region for stair blue noise sampling. Ds-wave sampling maps such unrepresentable high frequency content to noise with a profile closer to that for step blue noise, while preserving the same $\nu_0$, and thus noise levels for low frequency content, as stair blue noise. Although noise is introduced into a larger range of higher frequencies, this noise is much less objectionable than the patterns introduced by stair blue noise. We illustrate this further in Figures~\ref{fig:CosSampling} and~\ref{fig:ComparisonPatterns} for images with repeated structures of several frequencies. The cleaner reconstructions of repeated structures of lower frequencies (top rows) provided by stair blue noise sampling come at the expense of mapping repeated structures of frequencies higher than the representable frequency to colored noise. This manifests itself as secondary patterns and higher levels of noise in the sampled images (bottom rows). Ds-wave sampling leads to as white as possible noise for these cases, combining advantages of step and stair blue noise sampling.

\subsubsection*{Ds-wave and Other Patterns} 

We compare ds-wave sampling to further patterns commonly used for anti-aliasing in Figure~\ref{fig:ComparisonOthers}. Dart throwing results in a relatively small $m$ and hence does not lead to objectionable aliasing artifacts. However, it also leads to low frequency noise in sampled images, as is apparent for the zone plate image. For a wider range of cleaner low frequencies for sampled images (i.e. a larger low frequency region in power spectrum), CCCVT centroids~\cite{Balzer09Capacity} can be utilized. This results, however, in higher peak values, and thus more pronounced aliasing artifacts. 

An even larger low frequency region is possible with single-peak blue noise sampling~\cite{Heck13Blue}, as illustrated in Figure~\ref{fig:ComparisonOthers}, middle. Note that single-peak blue noise is not exactly zero at low frequencies due to the introduced Gaussian at around the transition from low to high frequencies, while ds-wave sampling has zero energy in the low frequency region. We set $\nu_0$ such that both single-peak and ds-wave reach $1$ the first time at approximately the same $\nu$. At this size of the low frequency region, the peak has a high value and aliasing becomes apparent as secondary patterns in the zone plate image in Figure~\ref{fig:ComparisonOthers}, middle, and the sampled high frequency repeated stripe patterns (bottom rows) in Figure~\ref{fig:ComparisonPatterns}. Our ds-wave sampling at $\nu_0 = 0.85$ and $e_0 = 0$ has the same size of the low frequency region, but with a lower maximum, and hence leads to lower noise and aliasing, as visible in the same figures.  

The maximum $m$ and hence aliasing artifacts can be further reduced by introducing low frequency noise. With $e_0 = 0.1$, we get a slightly smaller $m$ than dart throwing, while still having cleaner and a larger range of low frequencies than dart throwing as shown in Figure~\ref{fig:ComparisonOthers}.

\subsubsection*{Multiple Samples per Pixel} 

One way of reducing noise is increasing number of samples per pixel. However, if $P$ contains high peaks, for finite spp, there will always be secondary patterns due to aliasing when sampling image content of certain frequencies. To see this, we start by noting that the error after resampling to a regular grid is given by $\lvert K(\bm{\nu}) \rvert^2 E(\bm{\nu}) = \lvert K(\bm{\nu}) \rvert^2 \frac{1}{\lambda} \left[ P_t * (U + 1) \right] (\bm{\nu})$, as derived in Section~\ref{sec:errorAndSpectrum}. Increasing spp means we are keeping $K$ the same, and expanding $P$ (as it is related to $g$ with a Fourier transform, which compresses for larger number of points due to smaller distances among them, please see Section~\ref{sec:background}). If $P$ has peaks, they will thus be shifted to higher frequencies and be smoothed as a result of this expansion. If a sampled image has local structures of those frequencies, due to the convolution in the definition of $E(\bm{\nu})$, these peaks will then be shifted to lower frequencies that are captured by the filter $K$. Hence, similar but smoothed artifacts in the form of visible secondary patterns will appear in the final reconstructed image.

This is illustrated in Figure~\ref{fig:ComparisonsSpp} for the cosine function in Figure~\ref{fig:CosSampling} with $\nu_c = 0.85$. We use $4$ spp for the top row, and $16$ spp for the bottom row. Note that as we always normalize frequencies by the number of sampling points, the absolute frequency shifts with the spp. For both cases, increasing spp does not help to reduce the visible secondary structures due to aliasing. In fact, higher spp might lead to perceptually more apparent secondary structures, e.g. for $16$ spp in Figure~\ref{fig:ComparisonsSpp}.

\subsubsection*{Artifacts on Rendered Images}
We illustrate such aliasing artifacts for practical rendered scenes in Figure~\ref{fig:renderingResults}. For this figure, we sample each of the dimensions for the light transport, except the image plane, densely. Hence, the spp reported corresponds to the image plane samples. Computing each of those image plane samples is thus a costly operation involving a numerical integration for all other dimensions. We show a reference image first, then the result of stair blue noise on a smaller image with $1$ or $2$ spp on the left, and the corresponding result with ds-wave sampling on the right. Note that we intentionally did not resample the rendered small images with nearest neighbor sampling to illustrate that applying standard filters on the images with aliasing artifacts does not alleviate the aliasing artifacts due to colored noise. For certain scenes such as the top image, increasing the spp from $1$ to $2$ makes the aliasing artifacts more apparent. In general, we observed that ds-wave sampling makes the most difference for directional repeated structures as exemplified in the figure. 




\subsubsection*{Running Time} 

The formulation of the optimization problems with linear and quadratic programming allows us to use efficient and robust solvers. For total variation energy (linear programming), it takes $1$-$5$ minutes for the solver to converge on a PC with Intel(R) Xeon(R) CPU ES-2680 v3 @ $2.5$ GHz, with the running time increasing for larger $\nu_0$. As this optimization is done once and offline, the main computational complexity comes from the point distribution generation procedure~\cite{Heck13Blue}, which takes about one minute to converge for $4096$ points.


 

%% file: TexFiles/Conclusions.tex
\section{Conclusions and Future Work}

We presented a theoretical and practical framework for analyzing aliasing, and generating sampling patterns with optimized properties for anti-aliasing via formulating the problem of generating realizable spectra as variational optimization. The resulting patterns lead to practical improvements in reducing aliasing artifacts due to colored noise, and the proposed theoretical framework allows us to explore and revise optimality measures used for anti-aliasing. We see many interesting uses of this framework for future research, some of which we summarize below.

\subsubsection*{Sampling for Integration}

Although we focused on anti-aliasing when reconstructing images in the scope of this paper, a very promising direction is to optimize power spectra for reducing error in numerical integration. Recent works~\cite{Oztireli16Integration,Pilleboue15Variance} have proved that the dependence of error on power spectrum is given by $\frac{1}{\lambda} \int_{-\infty}^{\infty}  P_t(\bm{\nu}) (U(\bm{\nu}) + 1)  d \bm{\nu}$, as we also discussed in Section~\ref{sec:errorAndSpectrum}. Minimizing this error will turn into a linear programming problem when formulated as variational optimization, similar to Equation~\ref{eq:continuousOptimizationProblem}. Once characteristics of integrands are determined, we can get specialized optimal spectra as well. 

\subsubsection*{Adaptive Anti-aliasing Patterns}

Similar to previous works, we considered non-adaptive anti-aliasing patterns, with no information on the actual image to be represented. Recent works~\cite{Roveri17General} show that sampling patterns with adaptive second order product densities can lead to significant accuracy improvements for image representation and processing. By combining our optimization framework with locally adaptive point distribution synthesis algorithms~\cite{Roveri17General}, we can obtain optimal adaptive sampling patterns for image reconstruction.

\subsubsection*{Exploration of Second Order Characteristics}

Previous works explore the space of valid second order characteristics either via analysis of available point patterns~\cite{Oztireli12Analysis}, or parametrized families of power spectra~\cite{Heck13Blue,Kailkhura16Stair}. Our framework can be used to explore this space without such constraints. As an example, we showed that optimal maximal values for power spectra for given $\nu_0$'s can be obtained (Figure~\ref{fig:OptimumV0M0Graph}). Similar results can be derived for other applications such as geometry sampling, physically-based simulations, or natural distributions.

\begin{figure*}[t!]
  \centering
  \includegraphics[width=0.99\linewidth]{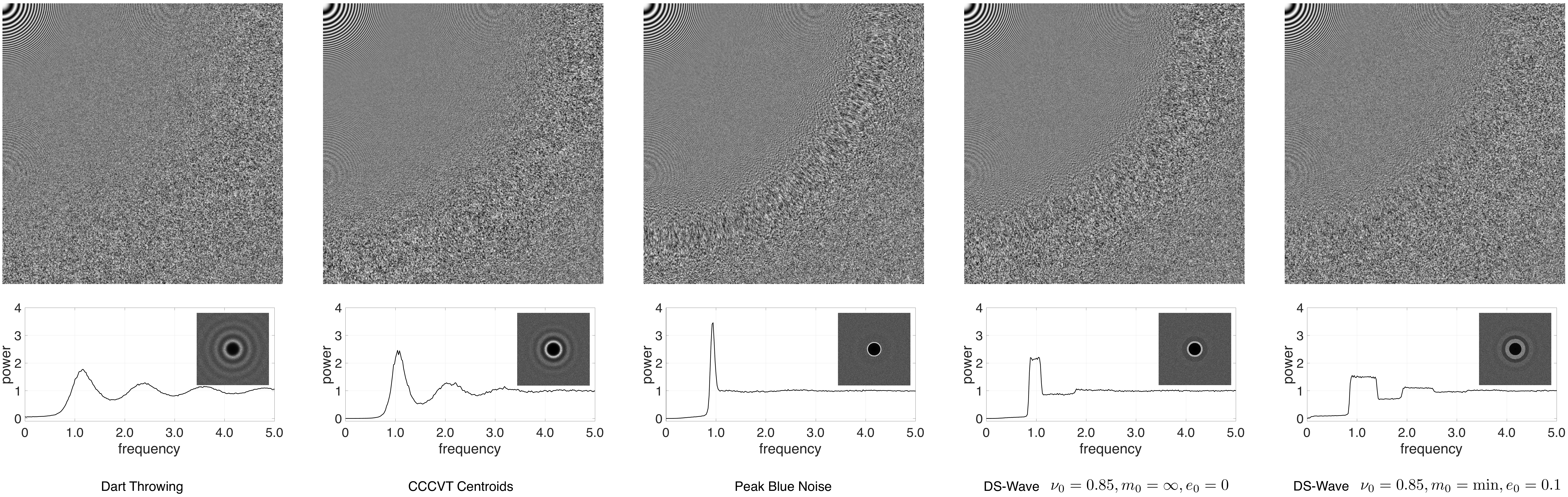}
 \caption{Reconstructed zone plate images and estimated power spectra for different sampling patterns. Ds-wave sampling can get the same effective range of the low frequency region as single-peak blue noise~\cite{Heck13Blue} with significantly less aliasing artifacts, which can be reduced even further by introducing low frequency noise with $e_0 = 0.1$.
  \label{fig:ComparisonOthers}}\vspace{-0mm}
\end{figure*}

\subsubsection*{Higher Dimensional Sampling}

An interesting aspect of the optimization problem in Equation~\ref{eq:continuousOptimizationProblem} is that it depends on the dimensionality due to the Hankel transform. We will thus get \emph{different spectra for different dimensions}, as Hankel transform takes a different form for different dimensions. It will be interesting to explore optimal sampling patterns for higher dimensions, e.g. in the context of rendering where the integrands can be very high dimensional.

\begin{figure}[t!]
  \centering
  \includegraphics[width=0.99\linewidth]{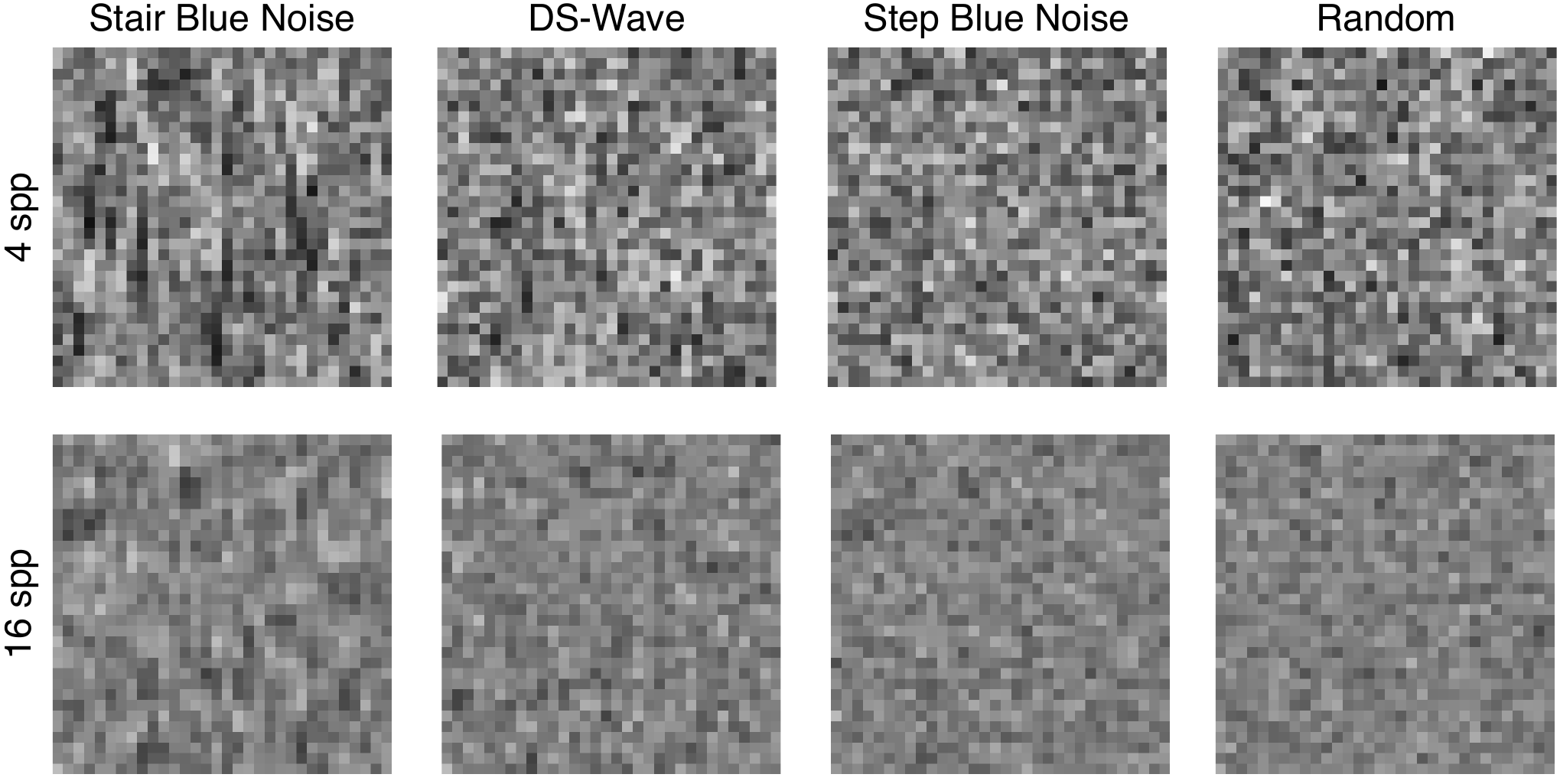}
 \caption{Increasing number of samples per pixel does not fundamentally solve the aliasing problem. We illustrate this when sampling the cosine function in Figure~\ref{fig:CosSampling} with $\nu_c = 0.85$ with stair blue noise sampling ($\nu_0 = 0.80$) for $4$ and $16$ spp. In contrast, ds-wave ($\nu_0 = 0.80$, $m_0 = \infty$), step blue noise, and random sampling do not lead to secondary patterns but to incoherent noise.
  \label{fig:ComparisonsSpp}}\vspace{-0mm}
\end{figure}

\subsubsection*{Synthesis of Point Patterns}

Our approach essentially formulates point pattern generation as a two-step procedure, where we first optimize for a power spectrum, and then generate point distributions with that power spectrum.  It is an ongoing research to synthesize point distributions with given statistics. Although the PCF based synthesis algorithm we use~\cite{Heck13Blue}, and others we tested~\cite{Oztireli12Analysis,Kailkhura16Stair} give very accurate results, all have a hard time to synthesize highly regular point sets (e.g. point distributions with $\nu_0 = 0.95$, $e_0 = 0$, and $m_0 = \mbox{min}$ in the supplementary material), as also observed in earlier works~\cite{Oztireli12Analysis}. As the synthesis algorithms evolve, the proposed formulation can be tuned further for the particular synthesis algorithm considered. For example, for a PCF based matching algorithm, the runtime can be reduced by considering a limited range for the PCF, which is possible if PCF is constant outside that range. This can be explicitly imposed as a constraint in our framework.


%% file: TexFiles/Appendix.tex

\section{Derivation of PCF as a Distribution}
\label{sec:appDerivationOfPCF}

Campbell's theorem~\cite{Illian08Statistical} gives sums of functions at sample points as integrals of those functions. For our case with the toroidal unit domain $\mathcal{V}$, we can write the theorem for first and second order product densities as
\begin{equation}
\mathbb{E}_{\mathcal{P}} \left[ \sum_j{t(\mathbf{x}_j)} \right] = \int_{\mathcal{V}}{t(\mathbf{x}) \varrho^{(1)} (\mathbf{x}) d\mathbf{x}}, 
\label{eq:campbellOneVariable}
\end{equation}
\begin{equation}
\mathbb{E}_{\mathcal{P}} \left[ \sum_{j \neq k}{t(\mathbf{x}_j, \mathbf{x}_k)} \right] = \int_{\mathcal{V} \times \mathcal{V}}{t(\mathbf{x}, \mathbf{y}) \varrho^{(2)} (\mathbf{x}, \mathbf{y}) d\mathbf{x} d\mathbf{y}}, 
\label{eq:campbellTwoVariables}
\end{equation}
provided some technical conditions are satisfied for the point process $\mathcal{P}$ and the function $t$~\cite{Illian08Statistical}. For stationary point processes, these simplify to
\begin{equation}
\mathbb{E}_{\mathcal{P}} \left[ \sum_j{t(\mathbf{x}_j)} \right] = \lambda \int_{\mathcal{V}}{t(\mathbf{x}) d\mathbf{x}},
\label{eq:campbellOneVariableStationary}
\end{equation}
\begin{equation}
\mathbb{E}_{\mathcal{P}} \left[ \sum_{j \neq k}{t(\mathbf{x}_j, \mathbf{x}_k)} \right] = \lambda^2 \int_{\mathcal{V} \times \mathcal{V}}{t(\mathbf{x}, \mathbf{y}) g(\mathbf{x} - \mathbf{y}) d\mathbf{x} d\mathbf{y}}.
\label{eq:campbellTwoVariablesStationary}
\end{equation}
Substituting $\delta(\mathbf{r} - (\mathbf{x}_j - \mathbf{x}_k)) = \delta(\mathbf{r} - \mathbf{r}_{jk})$ for $t(\mathbf{x}_j, \mathbf{x}_k)$ in Equation~\ref{eq:campbellTwoVariablesStationary}, we get
\begin{equation}
\begin{aligned}
\mathbb{E}_{\mathcal{P}} \left[ \sum_{j \neq k}{\delta(\mathbf{r} - (\mathbf{x}_j - \mathbf{x}_k))} \right] &= \lambda^2 \int_{\mathcal{V} \times \mathcal{V}}{ \delta(\mathbf{r} - (\mathbf{x} - \mathbf{y})) g(\mathbf{x} - \mathbf{y}) d\mathbf{x} d\mathbf{y}} \\
&= \lambda^2 g(\mathbf{r}),
\label{eq:pcfAsDistribution}
\end{aligned}
\end{equation}
proving that $g(\mathbf{r})$ can be estimated as the distribution of difference vectors.


\begin{figure}[t!]
  \centering
  \includegraphics[width=0.8\linewidth]{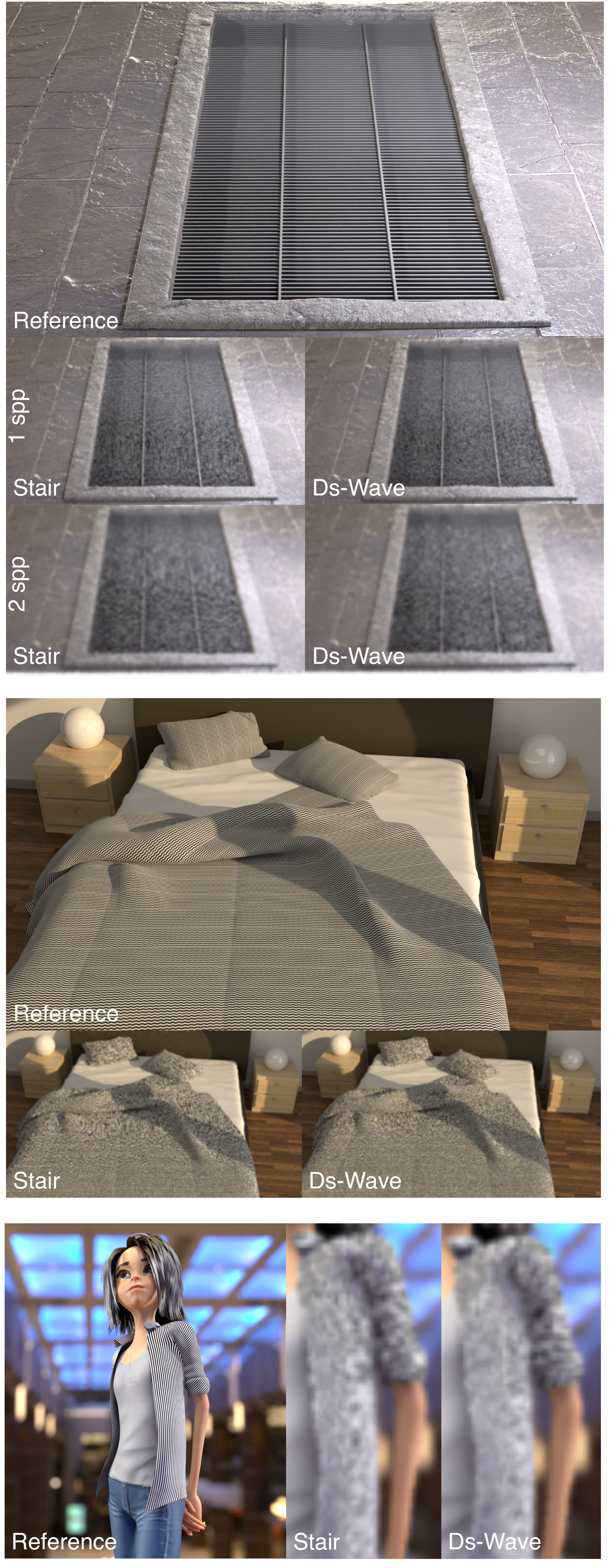}
 \caption{For each image, a reference rendering, and rendered images with $1$ or $2$ spp with stair blue noise, and ds-wave sampling are shown.
  \label{fig:renderingResults}}\vspace{-0mm}
\end{figure}

\section{Relation between PCF and Power Spectrum}
\label{sec:appRelationPCFAndPS}

The Fourier transform $G$ of PCF $g$ can be derived starting from the expression in Equation~\ref{eq:pcfAsDistribution} as follows
\begin{equation}
\begin{aligned}
G(\bm{\nu}) &= \mathscr{F}[g(\mathbf{r})](\bm{\nu}) = \frac{1}{\lambda^2} \mathbb{E}_{\mathcal{P}} \left[ \sum_{j \neq k}{ \mathscr{F} [ \delta(\mathbf{r} - (\mathbf{x}_j - \mathbf{x}_k))](\bm{\nu}) } \right] \\
&= \frac{1}{\lambda^2} \mathbb{E}_{\mathcal{P}} \left[ \sum_{j \neq k}{ e^{-2\pi i \bm{\nu}^T \mathbf{r}_{jk} } } \right].
\label{eq:fourierTransformOfPCFAsDistribution}
\end{aligned}
\end{equation}
We can further derive the following expression using Campbell's theorem for stationary point processes (Equation~\ref{eq:campbellOneVariableStationary})
\begin{equation}
\mathbb{E}_{\mathcal{P}} \sum_j 1 =  \lambda \int_{\mathcal{V}}{ d\mathbf{x}} = \lambda.
\label{eq:expectedValueOfOne}
\end{equation}
Summing these, we get
\begin{equation}
\begin{aligned}
\lambda G(\bm{\nu}) + 1 &= \frac{1}{\lambda} \mathbb{E}_{\mathcal{P}} \left[ \sum_{j \neq k}{ e^{-2\pi i \bm{\nu}^T \mathbf{r}_{jk}}} \right] + \frac{1}{\lambda}  \mathbb{E}_{\mathcal{P}} \sum_j 1 \\
&= \frac{1}{\lambda} \mathbb{E}_{\mathcal{P}} \left[ \sum_{jk}{ e^{-2\pi i \bm{\nu}^T \mathbf{r}_{jk}}} \right] = P(\bm{\nu})
\label{eq:pcfAndPSDerivation}
\end{aligned}
\end{equation}

\section{Derivation of Error Spectrum}
\label{sec:derivationPSError}

We start by rewriting the form of the error in Equation~\ref{eq:errorSpectrumExpansion}
\begin{equation}
\begin{aligned}
E &= \frac{1}{\lambda^2}\mathbb{E}_{\mathcal{P}} \left[ |S * T|^2 \right] + |T|^2
  - \frac{2}{\lambda} \Re \left\{ \left( \mathbb{E}_{\mathcal{P}} \left[ S\right] * T \right) \overline{T} \right\}.
\label{eq:errorSpectrumExpansionApp}
\end{aligned}
\end{equation}
As defined in Section~\ref{sec:background}, $s(\mathbf{x}) = \sum_j {\delta (\mathbf{x} - \mathbf{x}_j)}$ and thus its Fourier transform is $S(\bm{\nu}) = \sum_{j} e^{-2\pi i \bm{\nu}^T \mathbf{x}_{j}}$. Plugging this into the Campbell's theorem of first order (Equation~\ref{eq:campbellOneVariableStationary}) we get
\begin{equation}
\begin{aligned}
\mathbb{E}_{\mathcal{P}} \left[ S(\bm{\nu})\right] = \mathbb{E}_{\mathcal{P}} \left[ \sum_{j} e^{-2\pi i \bm{\nu}^T \mathbf{x}_{j}} \right]
= \lambda \int_{\mathcal{V}} e^{-2\pi i \bm{\nu}^T \mathbf{x}} d\mathbf{x} = \lambda \delta (\bm{\nu}).
\label{eq:expValS}
\end{aligned}
\end{equation}
The last term in Equation~\ref{eq:errorSpectrumExpansionApp} thus becomes $-2[\delta * T](\bm{\nu}) \overline{T(\bm{\nu})} = -2 \lvert T(\bm{\nu}) \rvert^2$. Calculating the first term in Equation~\ref{eq:errorSpectrumExpansionApp} is more involved due to the squared magnitude. We first expand this term with the definition of $S$ and utilizing properties of the Fourier transform
\begin{equation}
\begin{aligned}
\left\lvert \left[ S * T \right] (\bm{\nu}) \right\rvert^2 &= \left\lvert \sum_{j} e^{-2\pi i \bm{\nu}^T \mathbf{x}_{j}} * T(\bm{\nu}) \right\rvert^2 \\
&= \sum_{jk} \overline{ \left( e^{-2\pi i \bm{\nu}^T \mathbf{x}_{j}} * T(\bm{\nu}) \right) } \left( e^{-2\pi i \bm{\nu}^T \mathbf{x}_{k}} * T(\bm{\nu}) \right) \\
&=  \sum_{jk} t(\mathbf{x}_j) t(\mathbf{x}_k) e^{-2\pi i \bm{\nu}^T (\mathbf{x}_{k} - \mathbf{x}_{j})} \\
&= \sum_{j} t^2(\mathbf{x}_j) + \sum_{j \neq k} t(\mathbf{x}_j) t(\mathbf{x}_k) e^{-2\pi i \bm{\nu}^T (\mathbf{x}_{k} - \mathbf{x}_{j})},
\label{eq:expValSConvT1}
\end{aligned}
\end{equation}
where we used the notation $a(\bm{\nu}) * b(\bm{\nu})$ for $[a * b](\bm{\nu})$, and the equivalence $e^{-2\pi i \bm{\nu}^T \mathbf{x}_{j}} * T(\bm{\nu}) = \mathscr{F}[ \delta(\mathbf{x} - \mathbf{x}_j) t(\mathbf{x}) ] = \mathscr{F}[ \delta(\mathbf{x} - \mathbf{x}_j) t(\mathbf{x}_j) ] = e^{-2\pi i \bm{\nu}^T \mathbf{x}_{j}} t(\mathbf{x}_j)$. The expected value of the first term on the last line can be computed with Campbell's theorem of first order (Equation~\ref{eq:campbellOneVariableStationary}) as $\mathbb{E}_{\mathcal{P}} \left[ \sum_{j} t^2(\mathbf{x}_j) \right]  = \lambda \int_{\mathcal{V}} t^2(\mathbf{x}) d\mathbf{x}$. The second term involves a double sum, and the expected value can thus be computed by utilizing Equation~\ref{eq:campbellTwoVariablesStationary} as
\begin{equation}
\begin{aligned}
&\mathbb{E}_{\mathcal{P}} \left[ \sum_{j \neq k} t(\mathbf{x}_j) t(\mathbf{x}_k) e^{-2\pi i \bm{\nu}^T (\mathbf{x}_{k} - \mathbf{x}_{j})} \right] \\
&= \lambda^2 \int_{\mathcal{V} \times \mathcal{V}} t(\mathbf{x}) t(\mathbf{y}) e^{-2\pi i \bm{\nu}^T (\mathbf{x} - \mathbf{y})} g(\mathbf{x} - \mathbf{y}) d\mathbf{x} d\mathbf{y} \\
&= \lambda^2 \int_{\mathcal{V}} a_t(\mathbf{r}) e^{-2\pi i \bm{\nu}^T \mathbf{r}} g(\mathbf{r}) d\mathbf{r} \\
&= \lambda^2 \mathscr{F} \left[ a_t g \right](\bm{\nu}) = \lambda^2 \left[ \left\lvert T \right\rvert^2 * G \right] (\bm{\nu}),
\label{eq:expValSConvT2}
\end{aligned}
\end{equation}
with $a_t(\mathbf{r})$ denoting the autocorrelation of $t(\mathbf{x})$, and we use the relation $\mathscr{F}[a_t](\bm{\nu}) = |T(\bm{\nu})|^2$, and the multiplication theorem of Fourier transform. Substituting the expression for $G$ (Equation~\ref{eq:powerSpectrumAndPCFInTermsOfU}), this can also be written in terms of $U$ as $\lambda^2 \left[ \lvert T \rvert^2 * (U / \lambda + \delta) \right] (\bm{\nu}) = \lambda \left[ \lvert T \rvert^2 * U \right] (\bm{\nu}) + \lambda^2 \lvert T(\bm{\nu}) \rvert^2$. Summing the two terms in Equation~\ref{eq:expValSConvT1}, we thus get
\begin{equation}
\begin{aligned}
\mathbb{E}_{\mathcal{P}} \left[ \left\lvert \left[ S * T \right] (\bm{\nu}) \right\rvert^2 \right] &= \lambda \int_{\mathcal{V}} t^2(\mathbf{x}) d\mathbf{x} + \lambda \left[ \lvert T \rvert^2 * U \right] (\bm{\nu}) + \lambda^2 \lvert T(\bm{\nu}) \rvert^2.
\label{eq:expValSConvT3}
\end{aligned}
\end{equation}
Finally, we sum all the terms in Equation~\ref{eq:errorSpectrumExpansionApp}
\begin{equation}
\begin{aligned}
E(\bm{\nu}) &= \frac{1}{\lambda^2} \left( \lambda \int_{\mathcal{V}} t^2(\mathbf{x}) d\mathbf{x} + \lambda \left[ \lvert T \rvert^2 * U \right] (\bm{\nu}) + \lambda^2 \lvert T(\bm{\nu}) \rvert^2 \right) \\
&+ \lvert T(\bm{\nu}) \rvert^2 - 2\lvert T(\bm{\nu}) \rvert^2 \\
&= \frac{1}{\lambda} \left( \int_{\mathcal{V}} t^2(\mathbf{x}) d\mathbf{x} + \left[ \lvert T \rvert^2 * U \right] (\bm{\nu}) \right) \\
&= \frac{1}{\lambda} \left( \left[ \lvert T \rvert^2 * 1 \right] (\bm{\nu}) + \left[ \lvert T \rvert^2 * U \right] (\bm{\nu}) \right) \\
&= \frac{1}{\lambda} \left[ \lvert T \rvert^2 * (U + 1) \right] (\bm{\nu}).
\label{eq:expValSConvT4}
\end{aligned}
\end{equation}
